# Earth's Energy Imbalance and Implications

James Hansen, Makiko Sato, Pushker Kharecha

NASA Goddard Institute for Space Studies, New York, NY 10025, USA
Columbia University Earth Institute, New York, NY 10027, USA

Karina von Schuckmann

Centre National de la Recherche Scientifique, LOCEAN Paris, hosted by Ifremer, Brest, France

**Abstract.** Improving observations of ocean heat content show that Earth is absorbing more energy from the sun than it is radiating to space as heat, even during the recent solar minimum. The inferred planetary energy imbalance, $0.59 \pm 0.15$ W/m$^2$ during the 6-year period 2005-2010, confirms the dominant role of the human-made greenhouse effect in driving global climate change. Observed surface temperature change and ocean heat gain together constrain the net climate forcing and ocean mixing rates. We conclude that most climate models mix heat too efficiently into the deep ocean and as a result underestimate the negative forcing by human-made aerosols. Aerosol climate forcing today is inferred to be $1.6 \pm 0.3$ W/m$^2$, implying substantial aerosol indirect climate forcing via cloud changes. Continued failure to quantify the specific origins of this large forcing is untenable, as knowledge of changing aerosol effects is needed to understand future climate change. We conclude that recent slowdown of ocean heat uptake was caused by a delayed rebound effect from Mount Pinatubo aerosols and a deep prolonged solar minimum. Observed sea level rise during the Argo float era is readily accounted for by ice melt and ocean thermal expansion, but the ascendency of ice melt leads us to anticipate acceleration of the rate of sea level rise this decade.

Humanity is potentially vulnerable to global temperature change, as discussed in the Intergovernmental Panel on Climate Change (IPCC, 2001, 2007) reports and by innumerable authors. Although climate change is driven by many climate forcing agents and the climate system also exhibits unforced (chaotic) variability, it is now widely agreed that the strong global warming trend of recent decades is caused predominantly by human-made changes of atmospheric composition (IPCC, 2007).

The basic physics underlying this global warming, the greenhouse effect, is simple. An increase of gases such as $CO_2$ makes the atmosphere more opaque at infrared wavelengths. This added opacity causes the planet's heat radiation to space to arise from higher, colder levels in the atmosphere, thus reducing emission of heat energy to space. The temporary imbalance between the energy absorbed from the sun and heat emission to space, causes the planet to warm until planetary energy balance is restored.

The planetary energy imbalance caused by a change of atmospheric composition defines a climate forcing. Climate sensitivity, the eventual global temperature change per unit forcing, is known with good accuracy from Earth's paleoclimate history. However, two fundamental uncertainties limit our ability to predict global temperature change on decadal time scales.

First, although climate forcing by human-made greenhouse gases (GHGs) is known accurately, climate forcing caused by changing human-made aerosols is practically unmeasured. Aerosols are fine particles suspended in the air, such as dust, sulfates, and black soot (Ramanathan et al., 2001). Aerosol climate forcing is complex, because aerosols both reflect solar radiation to space (a cooling effect) and absorb solar radiation (a warming effect). In addition, atmospheric aerosols can alter cloud cover and cloud properties. Therefore, precise composition-specific measurements of aerosols and their effects on clouds are needed to assess the aerosol role in climate change.

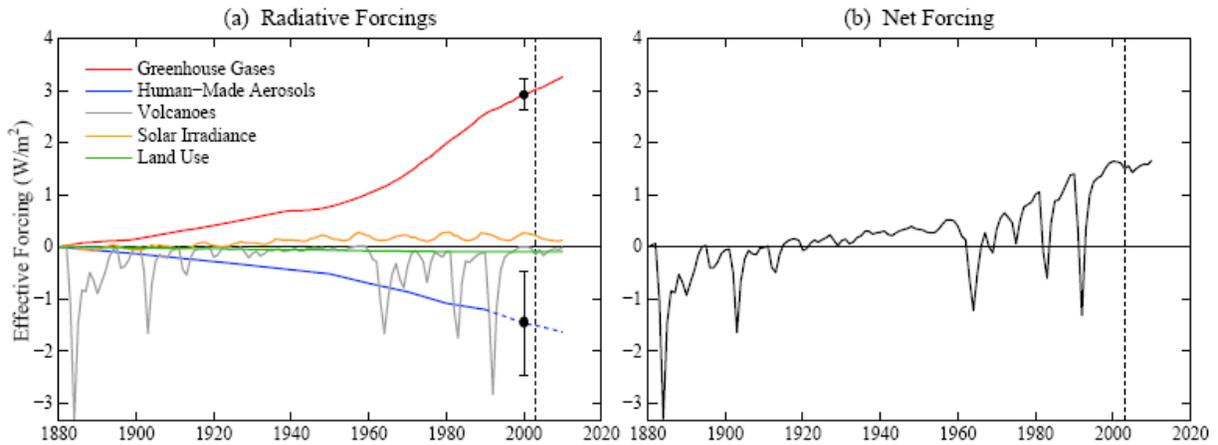

**Fig. 1.** Climate forcings employed in this paper. Forcings through 2003 (vertical line) are the same as used by Hansen et al. (2007), except the tropospheric aerosol forcing after 1990 is approximated as -0.5 times the GHG forcing. Aerosol forcing includes all aerosol effects, including indirect effects on clouds and snow albedo. GHGs include $O_3$ and stratospheric $H_2O$, in addition to well-mixed GHGs. These data are available at http://www.columbia.edu/~mhs119/EnergyImbalance/Imbalance.Fig01.txt

Second, the rate at which Earth's surface temperature approaches a new equilibrium in response to a climate forcing depends on how efficiently heat perturbations are mixed into the deeper ocean. Ocean mixing is complex and not necessarily simulated well by climate models. Empirical data on ocean heat uptake are improving rapidly, but still suffer limitations.

We summarize current understanding of this basic physics of global warming and note observations needed to narrow uncertainties. Appropriate measurements can quantify the major factors driving climate change, reveal how much additional global warming is already in the pipeline, and help define the reduction of climate forcing needed to stabilize climate.

## 1. Climate forcings

A climate forcing is an imposed perturbation of Earth's energy balance. Natural forcings include changes of solar irradiance and volcanic eruptions that inject aerosols to altitudes 10-30 km in the stratosphere, where they reside 1-2 years, reflecting sunlight and cooling Earth's surface. Principal human-made forcings are greenhouse gases and tropospheric aerosols, i.e., aerosols in Earth's lower atmosphere, mostly in the lowest few kilometers of the atmosphere.

A forcing, F, is measured in watts per square meter ($W/m^2$) averaged over the planet. For example, if the sun's brightness increases 1 percent the forcing is $F \sim 2.4$ $W/m^2$, because Earth absorbs about 240 $W/m^2$ of solar energy averaged over the planet's surface. If the $CO_2$ amount in the air is doubled[1], the forcing is $F \sim 4$ $W/m^2$. The opacity of a greenhouse gas as a function of wavelength is calculated via basic quantum physics and verified by laboratory measurements to an accuracy of a few percent. No climate model is needed to calculate the forcing due to changed greenhouse gas amount. It requires only summing over the planet the change of heat radiation to space, which depends on known atmospheric and surface properties.

---

[1] $CO_2$ climate forcing is approximately logarithmic, because its absorption bands saturate as $CO_2$ amount increases. An equation for climate forcing as a function of $CO_2$ amount is given in Table 1 of Hansen et al. (2000).

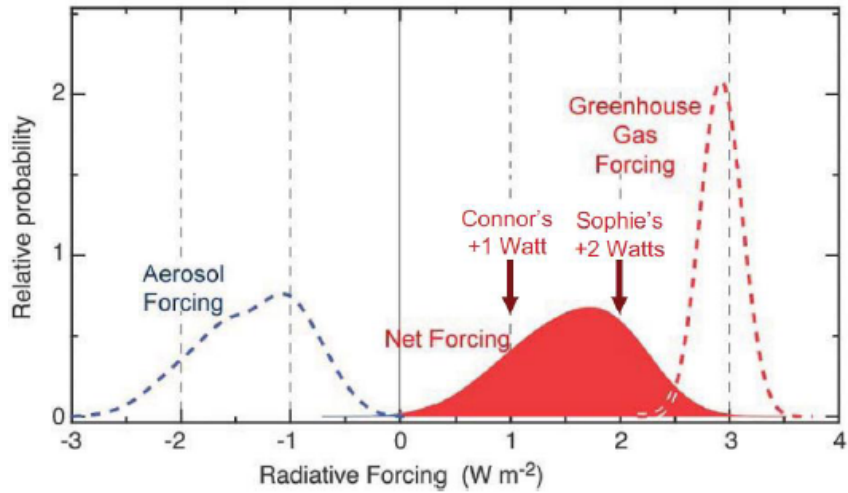

**Fig. 2.** Schematic diagram of human-made climate forcings by GHGs, aerosols, and their net effect in 2005, as presented in Figure 10 of Hansen (2009), which was adapted from IPCC (2007).

Fig. 1 shows climate forcings estimated by Hansen et al. (2007) and used by those authors for climate simulations with an atmosphere-ocean climate model. We will employ these forcings for simplified calculations of global temperature, demonstrating that a simple Green's function calculation, with negligible computation time, yields practically the same global temperature change as the complex climate model, provided that the global model's 'climate response function' has been defined. The response function specifies the fraction of the equilibrium (long-term) response achieved as a function of time following imposition of the forcing  The simplified calculations allow investigation of the consequences of errors in aerosol climate forcing and ocean heat uptake.

Fig. 2 is a slight adaptation of a figure in IPCC (2007), specifically the version used by Hansen (2009) to draw attention to the importance of the uncertainty in aerosol forcing. The arbitrary scale on the y-axis is meant to indicate that the GHG forcing is well known at about +3 W/m$^2$, while the aerosol forcing is only a heuristic estimate. The large uncertainty in the aerosol forcing implies that the net forcing is very uncertain, so uncertain that either value suggested by Hansen's grandchildren for the <u>net</u> forcing (Sophie's +2 W/m$^2$ or Connor's +1 W/m$^2$) could be correct.

The correct answer defines the terms of humanity's 'Faustian aerosol bargain' (Hansen and Lacis, 1990). Global warming has been limited, as aerosol cooling partially offsets GHG warming. But aerosols remain airborne only several days, so they must be pumped into the air faster and faster to keep pace with increasing long-lived GHGs. However, concern about health effects of particulate air pollution is likely to lead to eventual reduction of human-made aerosols. Thereupon the Faustian payment will come due.

If +2 W/m$^2$ net forcing is close to the truth (aerosol forcing  1 W/m$^2$), even a major effort to clean up aerosols, say reduction by half, increases the net forcing only 25 percent. But a net forcing of +1 W/m$^2$ (aerosol forcing  2 W/m$^2$) implies that reducing aerosols by half doubles the net climate forcing. Given global climate effects already being observed (IPCC, 2007), doubling the climate forcing suggests that humanity may face a grievous Faustian payment.

Most climate models in IPCC (2007) used aerosol forcing about -1 W/m$^2$. We will argue that this understates the true aerosol effect. But first we must discuss climate sensitivity.

## 2. Climate sensitivity and climate feedbacks

Climate sensitivity (S) is defined as the equilibrium global surface temperature change (ΔTeq) in response to a specified unit forcing after the planet has come back to energy balance,

$$S = \Delta Teq/F, \qquad (1)$$

i.e., climate sensitivity is the eventual global temperature change per unit forcing (F).

Climate sensitivity depends upon climate feedbacks, the many physical processes that come into play as climate changes in response to a forcing. Positive (amplifying) feedbacks increase the climate response, while negative (diminishing) feedbacks reduce the response.

Climate feedbacks do not come into play coincident with the forcing, but rather in response to climate change. Feedbacks operate by altering the amount of solar energy absorbed by the planet or the amount of heat radiated to space. It is assumed that, at least to a useful approximation, feedbacks affecting the global mean response are a function of global temperature change.

'Fast feedbacks' appear almost immediately in response to global temperature change. For example, as Earth becomes warmer the atmosphere holds more water vapor. Water vapor is an amplifying fast feedback, because water vapor is a powerful greenhouse gas. Other fast feedbacks include clouds, natural aerosols, snow cover and sea ice.

'Slow feedbacks' may lag global temperature change by decades, centuries, millennia, or longer time scales. Studies of paleoclimate, Earth's climate history, reveal that principal slow feedbacks are surface albedo (reflectivity, literally 'whiteness') and long-lived GHGs, and slow feedbacks are predominately amplifying feedbacks on millennial time scales (IPCC, 2007; Hansen et al., 2008; Hansen and Sato, 2011).

Our present paper concerns the past century. We can study this period making use of only the fast-feedback climate sensitivity. Slow feedback effects on long-lived greenhouse gases (GHGs) during the past century are implicitly included by using observed GHG amounts. Changes of Greenland and Antarctic ice sheet area during the past century are negligible.

Fast-feedback climate sensitivity has been estimated in innumerable climate model studies, most famously in the Charney et al. (1979) report that estimated equilibrium global warming of 3°C ± 1.5°C for doubled $CO_2$ (a forcing of 4 W/m$^2$), equivalent to 0.75°C ± 0.375°C per W/m$^2$. Subsequent model studies have not much altered this estimate or greatly reduced the error estimate, because of uncertainty as to whether all significant physical processes are included in the models and accurately represented. The range of model results in the most recent IPCC report was 2.1 – 4.4°C for doubled $CO_2$ (Randall et al., 2007).

Empirical assessment of the fast-feedback climate sensitivity can be extracted from glacial-interglacial climate oscillations, during which Earth was in quasi-equilibrium with slowly changing boundary forcings (Hansen and Sato, 2011). This assessment depends on knowledge of global temperature change and the GHG and surface albedo forcings, the latter depending mainly upon ice sheet size and thus upon sea level. Hansen and Sato (2011) use data for the past 800,000 years to conclude that the fast-feedback sensitivity is 0.75°C ± 0.125°C per W/m$^2$, which is equivalent to 3°C ± 0.5°C for doubled $CO_2$. This 1 σ error estimate is necessarily partly subjective. We employ fast-feedback climate sensitivity 0.75°C per W/m$^2$ in our present study.

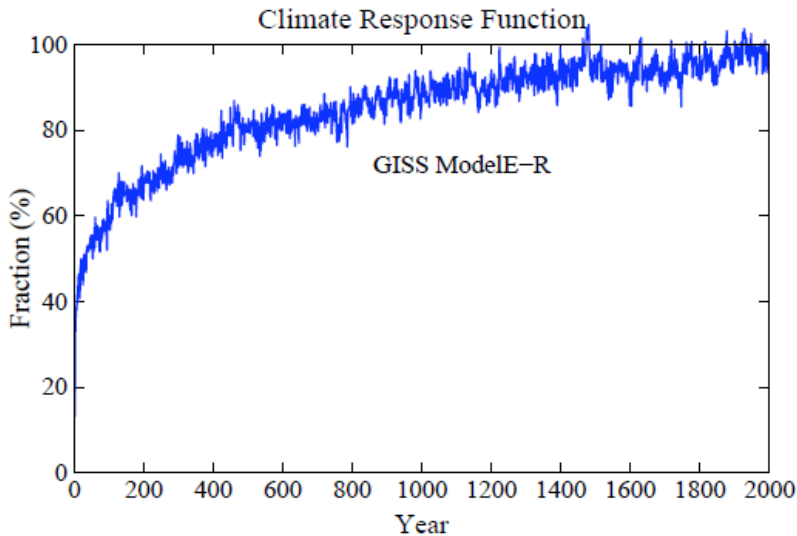

**Fig. 3.** Climate response function, R(t), i.e., the fraction of equilibrium surface temperature response for GISS climate model-ER, based on the 2000 year control run E3 (Hansen et al., 2007). Forcing was instant $CO_2$ doubling with fixed ice sheets, vegetation distribution, and other long-lived GHGs.

### 3. Climate response function

Climate response to human and natural forcings can be simulated with complex global climate models, and, using such models, it has been shown that warming of the ocean in recent decades can be reproduced well (Barnett et al, 2005; Hansen et al., 2005; Pierce et al., 2006). Here we seek a simple general framework to examine and compare models and the real world in terms of fundamental quantities that elucidate the significance of the planet's energy imbalance.

Global surface temperature does not respond quickly to a climate forcing, the response being slowed by the thermal inertia of the climate system. The ocean provides most of the heat storage capacity, because approximately its upper 100 meters is rapidly mixed by wind stress and convection (mixing is deepest in winter at high latitudes, where mixing occasionally extends into the deep ocean). Thermal inertia of the ocean mixed layer, by itself, would lead to a surface temperature response time of about a decade, but exchange of water between the mixed layer and deeper ocean increases the surface temperature response time by an amount that depends on the rate of mixing and climate sensitivity (Hansen et al., 1985).

The lag of the climate response can be characterized by a climate response function, which is defined as the fraction of the fast-feedback equilibrium response to a climate forcing. This response function is obtained from the temporal response of surface temperature to an instantaneously applied forcing, for example a doubling of atmospheric $CO_2$. Fig. 3 shows the response function of GISS modelE-R, which is the GISS atmospheric model (Schmidt et al., 2006) coupled to the Russell ocean model (Russell et al., 1995). This model has been characterized in detail via its response to many forcings (Hansen et al., 2005b, 2007).

About 40 percent of the equilibrium response is obtained within five years. This quick response is due to the small effective inertia of continents, but warming over continents is limited by exchange of continental and marine air masses. Only 60 percent of the equilibrium response is achieved in a century. Nearly full response requires a millennium.

Below we argue that the real world response function is faster than that of modelE-R. We also suggest that most global climate models are similarly too sluggish in their response to a climate forcing and that this lethargy has important implications for predicted climate change. It would be useful if response functions as in Fig. 3 were computed for all climate models to aid climate analysis and intercomparisons. Also, as shown in the next section, the response function can be used for a large range of climate studies.

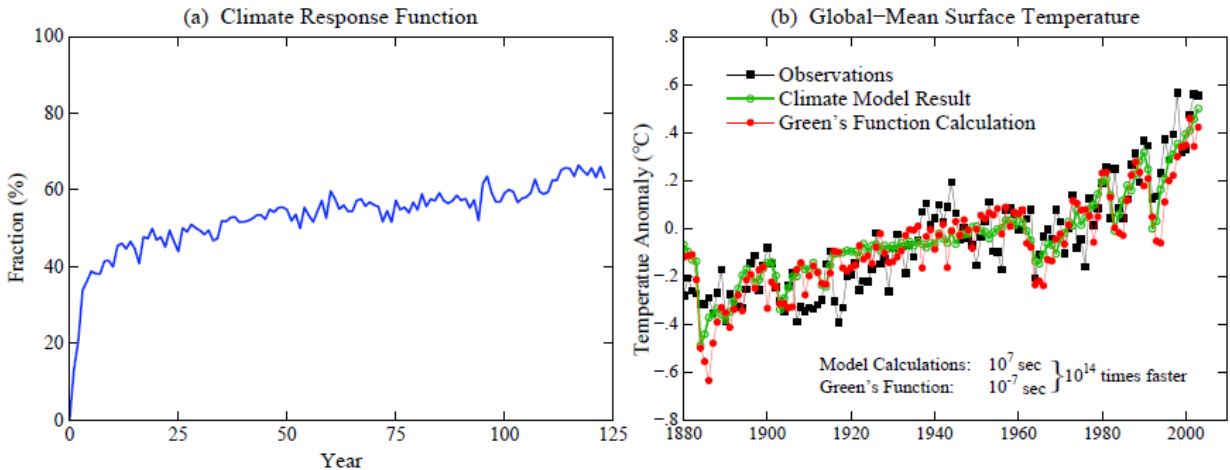

**Fig. 4.** (a) First 123 years of climate response function, from Fig. 3, (b) comparison of observed global temperature, mean result of 5-member ensemble of simulations with the GISS global climate modelE-R, and the simple Green's function calculation using the climate response function in Fig. 4a.

Held et al. (2010) show global temperature change obtained in 100-year simulations after instant $CO_2$ doubling for the Geophysical Fluid Dynamics Laboratory (GFDL) climate model, a model with equilibrium sensitivity 3.4°C for doubled $CO_2$. Held et al. (2010) and Winton et al. (2010) draw attention to and analyze two distinct time scales in the climate response, a quick partial climate response with characteristic time about 5 years and a slow warming on century time scales, which they term the 'recalcitrant' component of the climate response because it responds so sluggishly to change of the climate forcing. This decomposition provides useful insights that we will return to in our later discussion. The GISS modelE-R yields a similar response, as is more apparent with the higher temporal resolution of Fig. 4a.

Climate response time depends on climate sensitivity as well as on ocean mixing. The reason is that climate feedbacks come into play in response to temperature change, not in response to climate forcing. On a planet with no ocean or only a mixed layer ocean, the climate response time is proportional to climate sensitivity. However, with a realistic ocean that has exchange between the mixed layer and deeper ocean, the longer response time with higher sensitivity also allows more of the deep ocean heat capacity to come into play.

Hansen et al. (1985) show analytically, with ocean mixing approximated as a diffusive process, that the response time increases as the square of climate sensitivity. Thus a climate model or climate system with sensitivity 4°C for doubled $CO_2$ requires four times longer to approach equilibrium compared with a system having climate sensitivity 2°C for doubled $CO_2$.

The response function in Fig. 3 is derived from a climate model with sensitivity 3°C for doubled $CO_2$. When the response function of other models is evaluated, it would be most useful if the equilibrium climate sensitivity were also specified. Note that it is not necessary to run a climate model for millennia to determine the equilibrium response. The remaining planetary energy imbalance at any point in the model run defines the portion of the original forcing that has not yet been responded to, which permits an accurate estimate of the equilibrium response via an analytic expression (Equation 3, Section 8 below) or linear regression of the planetary energy imbalance against surface temperature change (Gregory et al., 2004).

## 4. Green's function

The climate response function, R(t), is a Green's function that allows calculation of global temperature change from an initial equilibrium state for any climate forcing history (Hansen, 2008),

$$T(t) = \int R(t) [dF/dt] \, dt . \qquad (2)$$

Fig. 4b shows results of this simple integration. R is the response function in Figs. 3 and 4a. F is the sum of the forcings in Fig. 1, and dF/dt is the annual increment of this forcing. The integration extends from 1880 to 2003, the period of the global climate model simulations of Hansen et al. (2007) illustrated in Fig. 4b.

The green curve in Fig. 4 is the mean result from five runs of the GISS atmosphere-ocean climate modelE-R (Hansen et al., 2007). Each of the runs in the ensemble used all the forcings in Fig. 1. Chaotic variability in the climate model ensemble is minimized by the 5-run mean.

The Green's function result (red curve in Fig. 4) has interannual variability, because the response function is based on a single climate model run that has unforced (chaotic) variability. Interannual variability in observations is larger than in the model, partly because the amplitude of Southern Oscillation (El Nino/La Nina) variability is unrealistically small in GISS modelE-R. Variability in the observed curve is also increased by the measurement error in observations, especially in the early part of the record.

Timing of chaotic oscillations in the response function (Fig. 3 and 4a) is accidental. Thus for Green's function calculations below we fit straight lines to the response function, eliminating the noise. Global temperature change calculated from Eq. (3) using the smoothed response function lacks realistic appearing year-to-year variability, but the smoothed response function provides a clearer correspondence with climate forcings. Except for these chaotic fluctuations, the results using the original and the smoothed response functions are very similar.

## 5. Alternative response functions

We believe, for several reasons, that the GISS modelE-R response function in Figs. 3 and 4a is slower than the climate response function of the real world. First, the ocean model mixes too rapidly into the deep Southern Ocean, as judged by comparison to observed transient tracers such as chlorofluorocarbons (CFCs) (Romanou and Marshall, private communication, paper in preparation). Second, the ocean thermocline at lower latitudes is driven too deep by excessive downward transport of heat, as judged by comparison with observed ocean temperature (Levitus and Boyer, 1994). Third, the model's low-order finite differencing scheme and parameterizations for diapycnal and mesoscale eddy mixing are excessively diffusive, as judged by comparison with relevant observations and LES (large eddy simulation) models (Canuto et al., 2010).

A substantial effort is underway to isolate the causes of excessive vertical mixing in the GISS ocean model (J. Marshall, private communication), including implementation of higher order finite differencing schemes, increased spatial resolution, replacement of small-scale mixing parameterizations with more physically-based methods (Canuto et al., 2010), and consideration of possible alternatives for the vertical advection scheme. These issues, however, are difficult and long-standing. Thus, for the time being, we estimate alternative climate response functions based on intuition tempered by evidence of the degree to which the model tracer transports differ from observations in the Southern Ocean and the deepening of the thermocline at lower latitudes.

The Russell ocean model, defining our 'slow' response function, achieves only 60 percent response after 100 years. As an upper limit for a "fast" response we choose 90 percent after 100 years (Fig. 5). It is unlikely that the ocean surface temperature responds faster than that, because the drive for transport of energy into the ocean is removed as the surface temperature approaches equilibrium and the planet achieves energy balance. We know from paleoclimate data that the lag of glacial-interglacial deep ocean temperature change, relative to surface temperature change, is not more than of the order of a millennium, so the energy source to the deep ocean must not be cut off too rapidly. Thus we are confident that the range from 60 to 90 percent encompasses the real world response at 100 years. As an intermediate response function we take 75 percent surface response at 100 years, in the middle of the range that is plausible for climate sensitivity 3°C for doubled $CO_2$.

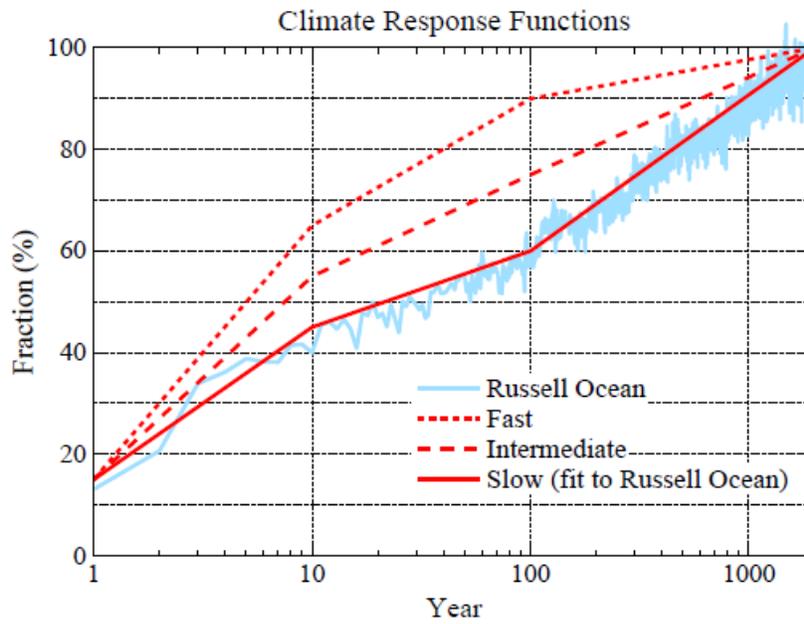

**Fig. 5.** Alternative climate response functions. The "slow" response is representative of many global climate models (see text).

The shape of the response function is dictated by the fact that the short-term response cannot be much larger than it is in the existing model (the slow response function). Indeed, the observed climate response to large volcanic eruptions, which produce a rapid (negative) forcing, suggests that the short-term model response is somewhat larger than in the real world. Useful volcanic tests, however, are limited to the small number of large eruptions occurring since the late 1800s (Robock, 2000; Hansen et al., 1996), with volcanic aerosol forcing uncertain by 25-50 percent. Almost invariably, an El Nino coincided with the period of predicted cooling, thus reducing the global cooling. Conceivably volcanic aerosols affect atmospheric temperatures in a way that induces El Ninos (Handler, 1986), but the possibility of such intricate dynamical effects should not affect deep ocean heat sequestration on longer time scales. Also, such an effect, if it exists, probably would not apply to other forcings, so it seems unwise to adjust the short-term climate response function based on this single empirical test.

## 6. Generality of slow response

We suspect that the slow response function of GISS modelE-R is common among many climate models reported in IPCC (2001, 2007) studies. WCRP (World Climate Research Program) requests modeling groups to perform several standard simulations, but existing tests do not include instantaneous forcing and long runs that would define the response function. However, Gokhan Danabasoglu provided us results of a 3000 year run of the NCAR (National Center for Atmospheric Research) CCSM3 model in response to instant $CO_2$ doubling. Tom Delworth provided us the global temperature history generated by the GFDL CM2.1 model, another of the principal IPCC models, the same model discussed above and by Held et al. (2010).

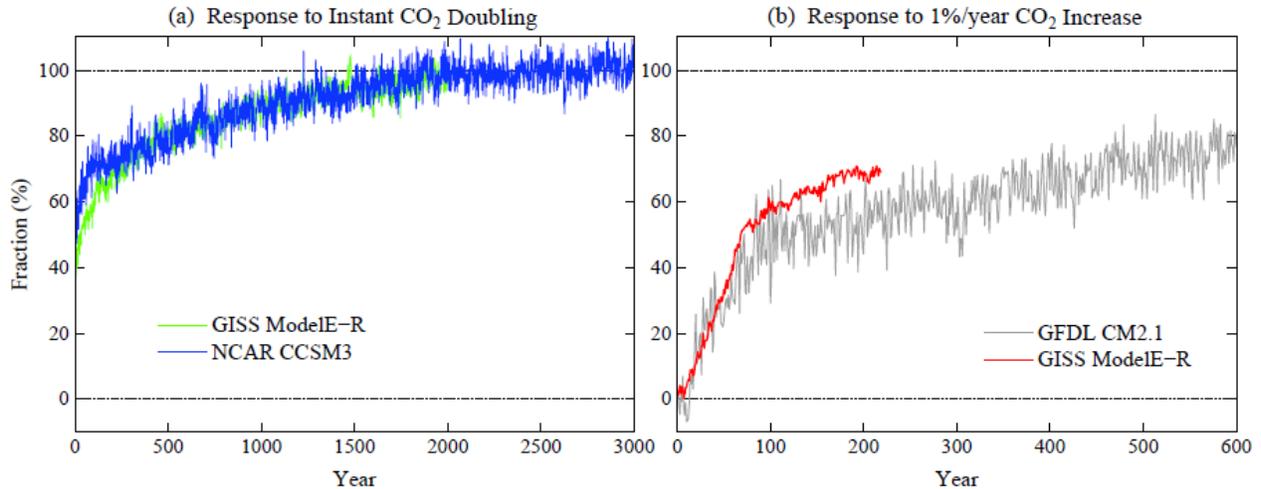

**Fig. 6.** (a) Climate response function of NCAR CCSM3 model for instant $CO_2$ doubling, directly comparable to GISS modelE-R result in Fig. 3, (b) response of GFDL CM2.1 and GISS modelE-R to 1%/year $CO_2$ increase until $CO_2$ doubling achieved (year 70) with $CO_2$ thereafter held constant.

Fig. 6a shows the response function of the NCAR CCSM3 model (Kiehl et al., 2006), which is directly comparable to the response of the GISS model in Fig. 3. Fig. 6b compares the GFDL and GISS model responses to a 1%/year $CO_2$ forcing. Equilibrium sensitivities of these specific NCAR and GFDL models are 2.5°C and 3.4°C, respectively, which compares to 3.0°C for the GISS model. It is clear from Fig. 6 that ocean mixing slows the surface temperature response about as much in the other two models as in the GISS model, with the differences being consistent with their moderate differences in equilibrium sensitivity. Data provided by J. Gregory (private communication, 2008) for a 1200-year run of the UK Hadley Centre model (Gordon et al., 2000) imply a still longer response time for that model, which is consistent with comparably efficient mixing of heat into the deep ocean, given the greater climate sensitivity of that model (about 10°C for quadrupled $CO_2$, which is a forcing ~8 W/m$^2$).

One plausible explanation for why many models have similarly slow response functions is common ancestry. The ocean component of many atmosphere-ocean climate models is the GFDL Bryan-Cox ocean model (Bryan, 1969; Cox, 1984). Common ancestry of the ocean sub-model is true for some of the principal models contributing to the IPCC (2001, 2007) climate studies, including (1) Parallel Climate Model (PCM), which uses the NCAR CCM3 atmosphere and land model with the Department of Energy Parallel Ocean Program ocean model (Washington et al., 2000), (2) GFDL R30 coupled climate model (Delworth et al., 2002), which uses version 1.1 of the Modular Ocean Model (Pacanowski et al., 1991), and (3) HadCM3 with atmosphere model described by Pope et al. (2000) and ocean model by Gordon et al. (2000).

Although models with independent ancestry exist, e.g., the isopycnal model of Bleck (2002), excessive mixing may arise from common difficulties, numerical and phenomenological, in simulating ocean processes, e.g.: (1) the vertical column in the Southern Ocean is only marginally stable, so flaws in simulating the surface climate in that region can lead to excessive mixing into the deep ocean, as occurs in GISS modelE-R, (2) imprecision in numerical finite-difference calculations may cause artificial diffusion, (3) commonly used parameterizations for subgrid-scale mixing may be excessively diffusive, as suggested by Canuto et al. (2010).

Comparisons of many climate models by Forest et al. (2006) and Stott and Forest (2007) show that ocean mixing and heat uptake in the models discussed above are typical of the models used in IPCC climate studies. One conclusion of Forest et al. (2006) is that most models, if not all, mix heat into the deep ocean too efficiently, compared to observed rates of ocean warming.

## 7. Implication of excessive ocean mixing

If the models assessed in the IPCC reports mix heat downward more efficiently than the real world, it raises the question of why the models do a good job of simulating the magnitude of global warming over the past century. The likely answer becomes apparent upon realization that the surface temperature change depends upon three factors: (1) the net climate forcing, (2) the equilibrium climate sensitivity, and (3) the climate response function, which depends on the rate at which heat is transported into the deeper ocean (beneath the mixed layer).

The equilibrium climate sensitivity of most of the models is within or close to the range 3 ± 0.5°C for doubled $CO_2$ dictated by paleoclimate data (Hansen and Sato, 2011). Thus, if the climate response function of the models is too slow, yet they achieve the observed magnitude of global warming, then the models must employ a net climate forcing that is larger than the climate forcing in the real world.

Knutti (2008) suggests that the explanation might be provided by the fact that most of the models employ a (negative) aerosol forcing that is smaller in magnitude than the aerosol forcing estimated *a priori* by IPCC. Indeed, most IPCC climate models exclude indirect aerosol forcing, i.e., the effect of human-made aerosols on clouds. IPCC estimates the 2005 aerosol cloud albedo effect as -0.7 W/m$^2$, uncertain by about a factor of two. IPCC also suggests the likelihood of a comparable, but unquantified, negative forcing due to aerosol effects on cloud cover.

If understated aerosol forcing is the correct explanation, producing a too-large net forcing that compensates for ocean models that mix heat too efficiently, there are important implications. The IPCC GHG forcing is about 3 W/m$^2$ in 2005 (2.66 W/m$^2$ from long-lived GHGs and 0.35 W/m$^2$ from tropospheric $O_3$. The typical aerosol forcing in the climate models is about -1 W/m$^2$, leaving a net forcing of about 2 W/m$^2$.

If the negative aerosol forcing is understated by as much as 0.7 W/m$^2$, it means that aerosols have been counteracting half or more of the GHG forcing. In that event, humanity has made itself a Faustian bargain more dangerous than commonly supposed.

## 8. Ambiguity between aerosols and ocean mixing

Uncertainties in aerosol forcing and ocean mixing (climate response function) imply that there is a family of solutions consistent with observed global warming. The range of acceptable solutions is explored in Fig. 7 via Green's function calculations that employ the three response functions of Fig. 5. The slow response function, based on GISS modelE-R, is typical of most IPCC models. The fast response function has a minimal rate of mixing into the deep ocean. The intermediate response function is a conjecture influenced by knowledge of excessive ocean mixing in GISS modelE-R.

Temporal variation of aerosol forcing is assumed to be proportional to the aerosol forcing in Fig. 1a. We seek the value of a factor ('constant') multiplying this aerosol forcing history that yields closest agreement with the observed temperature record. In a later section we will discuss uncertainties in the shape of the aerosol forcing history in Fig. 1a.

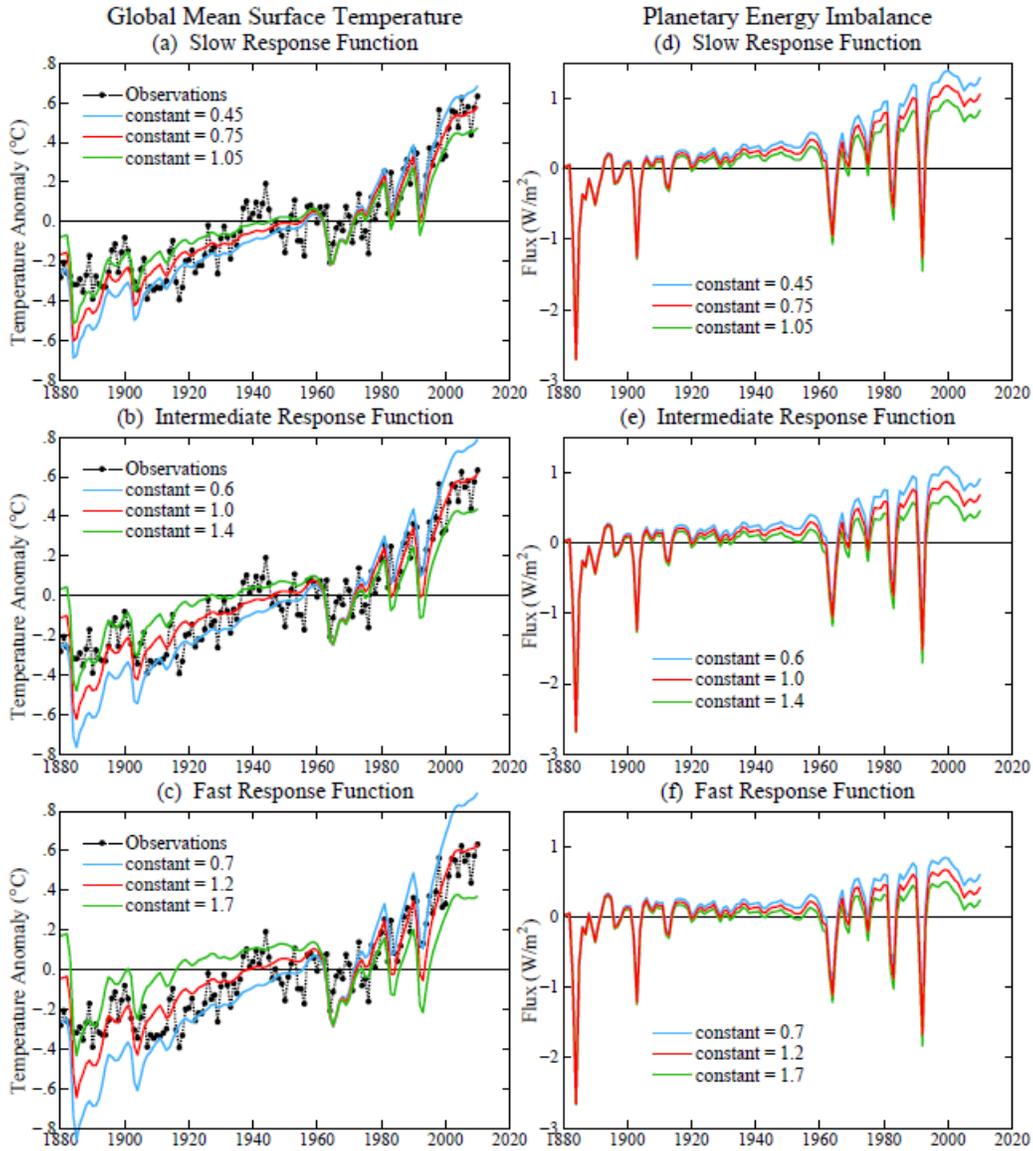

**Fig. 7.** Green's function calculation of surface temperature change and planetary energy imbalance. Three choices for climate response function are slow (top row, same as GISS modelE-R), intermediate (middle) and fast response (bottom). Factor 'constant' multiplies aerosol forcing of Fig. 1.

Green's function calculations were made for many values of constant. Values of constant providing best least-squares fit for the entire 1880-2010 period are 0.84, 1.05 and 1.20 for the slow, intermediate and fast response functions. This entire record may give too much weight to the late 1800s, when the volcanic aerosol optical depth from Krakatau and other volcanoes is very uncertain and the global temperature record is also least certain. Thus we also found the values of constant providing best fit for the period 1950-2010: 0.68, 0.97 and 1.16 for the slow, intermediate and fast response functions. The least-squares error curves are quite flat-bottomed, so intermediate values for constant: 0.75, 1 and 1.2 fit the observed temperature curve nearly as well for both periods as the values optimized for a single period.

Thus the aerosol forcing that provides best agreement with observed global temperature for the slow response function (deep ocean mixing) is ~1.2 W/m$^2$. The best fit aerosol forcings are ~1.6 and ~2.0 W/m$^2$ for the intermediate and fast response functions.

Given that two major uncertainties (aerosol forcing and ocean mixing) affect expected global warming, solution of the problem requires a second criterion, in addition to global temperature. The planetary energy imbalance (Hansen et al., 1997, 2005) is the fundamental relevant quantity, because it is a direct consequence of the net climate forcing.

Expected planetary energy imbalance for any given climate forcing scenario and ocean mixing (climate response function) follows from the above Green's function calculation:

$$\text{Planetary Energy Imbalance (t)} = F \times (\Delta T_{eq} - \Delta T)/\Delta T_{eq} = F \ \Delta T/S \ . \qquad (3)$$

This equation is simply a statement of the fact that the energy imbalance is the portion of the climate forcing that the planet's surface temperature has not yet responded to. $\Delta T_{eq}$, $\Delta T$ and F in this equation are all functions of time. $\Delta T_{eq}$, the equilibrium temperature change for the climate forcing that existed at time t, is the product of the climate forcing at time t and the fast-feedback climate sensitivity, $S \times F$, with $S \sim \frac{3}{4}$°C per W/m$^2$. $\Delta T$ is the global surface temperature at time t calculated with the Green's function (3). Planetary energy imbalance calculated from (4) agrees closely with global climate model simulations (Hansen et al., 2007).

Equation (3) with $\Delta T_{eq}$ and S defined to include only fast feedbacks is valid for time scales from decades to a century, a period short enough that the size of the ice sheets will not change significantly. The climate forcing, F, is defined to include all changes of long-lived gases including those that arise from slow carbon cycle feedbacks that affect atmospheric composition, such as those due to changing ocean temperature or melting permafrost.

Calculated planetary energy imbalance for the three ocean mixing rates, based on (4), are shown in the right half of Fig. 7. The slow response function, relevant to most climate models, has a planetary energy imbalance ~ 1 W/m$^2$ in the first decade of the 21$^{st}$ century. The fast response function has an average energy imbalance ~ 0.35 W/m$^2$ in that decade. The intermediate climate response function falls about half way between these extremes.

Discrimination among these alternatives requires observations of changing ocean heat content. Ocean heat data prior to 1970 are not sufficient to produce a useful global average, and data for most of the subsequent period are still plagued with instrumental error and poor spatial coverage, especially of the deep ocean and the Southern Hemisphere, as quantified in analyses and error estimates by Domingues et al. (2008) and Lyman and Johnson (2008).

Dramatic improvement in knowledge of Earth's energy imbalance is possible this decade as Argo float observations (Roemmich and Gilson, 2009) are improved and extended. If Argo data are complemented with adequate measurements of climate forcings, we will argue, it will be possible to assess the status of the global climate system, the magnitude of global warming in the pipeline, and the change of climate forcing that is required to stabilize climate.

## 9. Observed planetary energy imbalance

As ocean heat data improve, it is relevant to quantify smaller terms in the planet's energy budget. Levitus et al. (2005), Hansen et al. (2005a) and IPCC (2007) estimated past multi-decadal changes of small terms in Earth's energy imbalance. Recently improved data, including satellite measurements of ice, make it possible to tabulate many of these terms on an annual basis. These smaller terms may become increasingly important, especially if ice melting continues to increase, so continued satellite measurements are important.

Our units for Earth's energy imbalance (planetary heat storage) are W/m$^2$ averaged over Earth's entire surface (~ $5.1 \times 10^{14}$ m$^2$). Note that 1 watt-year for the full surface of Earth is approximately $1.61 \times 10^{22}$ J (joules).

## 9.1. Non-ocean terms in planetary energy imbalance

The variability of annual changes of heat content is large, but smoothing over several years allows trends to be seen. For consistency with the analysis of Argo data, we calculate 6-year moving trends of heat uptake for all of the terms in the planetary energy imbalance.

**Atmosphere.** The atmospheric term in the planet's energy imbalance is small, because the atmospheric heat capacity is small. Because the term is small, we obtain it simply as the product of the surface air temperature change in the analysis of Hansen et al. (2010), the mass of the atmosphere (~$5.13 \times 10^{18}$ kg), and its heat capacity $c_p \sim 1000$ J/kg/K. The fact that upper tropospheric temperature change tends to exceed surface temperature change is offset by stratospheric cooling that accompanies tropospheric warming. Based on simulated changes of atmospheric temperature profile (IPCC, 2007; Hansen et al., 2005), our use of surface temperature change to approximate mean atmospheric change modestly overstates heat content change. Fig. 8a shows the 6-year moving trend of atmospheric heat content change.

IPCC (2007) in their Figure 5.4 has atmospheric heat gain as the second largest non-ocean term in the planetary energy imbalance at $5 \times 10^{21}$ J for the period 1961-2003. Levitus et al. (2001) has it even larger at $6.6 \times 10^{21}$ J for the period 1955-1996. Our calculation yields $2.5 \times 10^{21}$ J for 1961-2003 and $2 \times 10^{21}$ J for 1955-1996. We could not find support for the larger values of IPCC (2007) and Levitus et al. (2001) in the references that they provided. The latent energy associated with increasing atmospheric water vapor in a warmer atmosphere is an order of magnitude too small to provide an explanation for the high estimates of atmospheric heat gain.

**Land.** We calculate transient ground heat uptake for the period 1880-2009 employing the standard one-dimensional heat conduction equation. Calculations went to a depth of 200 m, which is sufficient to capture heat content change on the century time scale. We used global average values of thermal diffusivity, mass density, and specific heat from Whittington et al. (2009). Temperature changes of the surface layer were driven by the global-land mean temperature change in the GISS data set (Hansen et al., 2010); a graph of global-land temperature is available at http://www.columbia.edu/~mhs119/Temperature/T_moreFigs/

Our calculated ground heat uptake is within the range of other estimates. For example, for the period 1901-2000 we obtain ~$12.6 \times 10^{21}$ J; Beltrami et al. (2002) give ~$15.9 \times 10^{21}$ J; Beltrami (2002) gives ~$13 \times 10^{21}$ J; Huang (2006) gives $10.3 \times 10^{21}$ J. Differences among these analyses are largely due to alternative approaches for deriving surface heat fluxes as well as alternative choices for the thermal parameters mentioned above (ours being based on Whittington et al., 2009). Our result is closest to that of Beltrami (2002), who derived land surface flux histories and heat gain directly from borehole temperature profiles (using a greater number of profiles than Beltrami et al., 2002).

**Ice on land.** We use gravity satellite measurements of mass changes of the Greenland and Antarctic ice sheets (Velicogna, 2009). For the period prior to gravity satellite data we extrapolate backward to smaller Greenland and Antarctic mass loss using a 10-year doubling time. Although Hansen and Sato (2011) showed that the satellite record is too short to well-define a curve for mass loss versus time, the choice to have mass loss decrease rapidly toward earlier times is consistent with a common glaciological assumption that the ice sheets were close to mass balance in the 1990s (Zwally et al., 2011). Our calculations for the energy associated with decreased ice mass assumes that the ice begins at -10°C and eventually reaches a mean temperature +15°C, but most of the energy is used in the phase change from ice to water.

Mass loss by small glaciers and ice caps is taken as the exponential fit to data in Fig. 1 of Meier et al. (2007) up to 2005 and as constant thereafter.

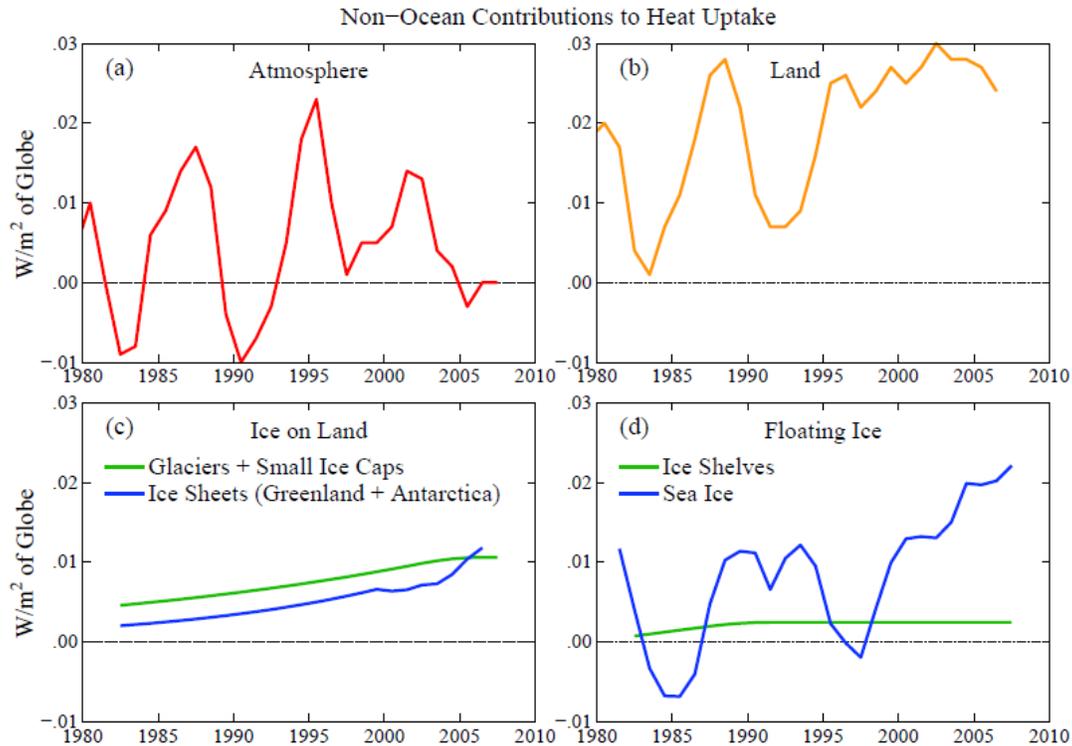

**Fig. 8.** Contributions to planetary energy imbalance by processes other than ocean heat uptake. Annual data is smoothed with moving 6-year linear trends.

**Floating ice.** Change of Arctic sea ice volume (Rothrock et al., 2008) is taken from (http://psc.apl.washington.edu/ArcticSeaiceVolume/IceVolume.php), data on the University of Washington Polar Science Center web site. Change of Antarctic sea ice area is from the National Snow and Ice Data Center (http://nsidc.org/data/seaice_index/archives/index.html), with thickness of Antarctic ice assumed to be one meter. Antarctic sea ice volume changes, and heat content changes, are small compared to the Arctic change.

We use the Shepherd et al. (2010) estimate for change of ice shelf volume, which yields a very small ice shelf contribution to planetary energy imbalance (Fig. 8d). Although Shepherd et al. (2010) have numerous ice shelves losing mass, with Larsen B losing an average of 100 cubic kilometers per year from 1998 to 2008, they estimate that the Filchner-Ronne, Ross, and Amery ice shelves are gaining mass at a combined rate of more than 350 cubic kilometers per year due to a small thickening of these large-area ice shelves.

**Summary.** Land warming (Fig. 8b) has been the largest of the non-ocean terms in the planetary energy imbalance over the past few decades. However, contributions from melting polar ice are growing rapidly. The very small value for ice shelves, based on Shepherd et al. (2010), seems questionable, depending very sensitively on estimated changes of the thickness of the large ice shelves. The largest ice shelves and the ice sheets could become major contributors to energy imbalance, if they begin to shed mass more rapidly. Because of the small value of the ice shelf term, we have neglected the lag between the time of ice shelf break-up and the time of melting, but this lag may become significant with major ice shelf breakup.

The sum of non-ocean contributions to the planetary energy imbalance is shown in Fig. 9a. This sum is still small, less than 0.1 W/m$^2$, but growing.

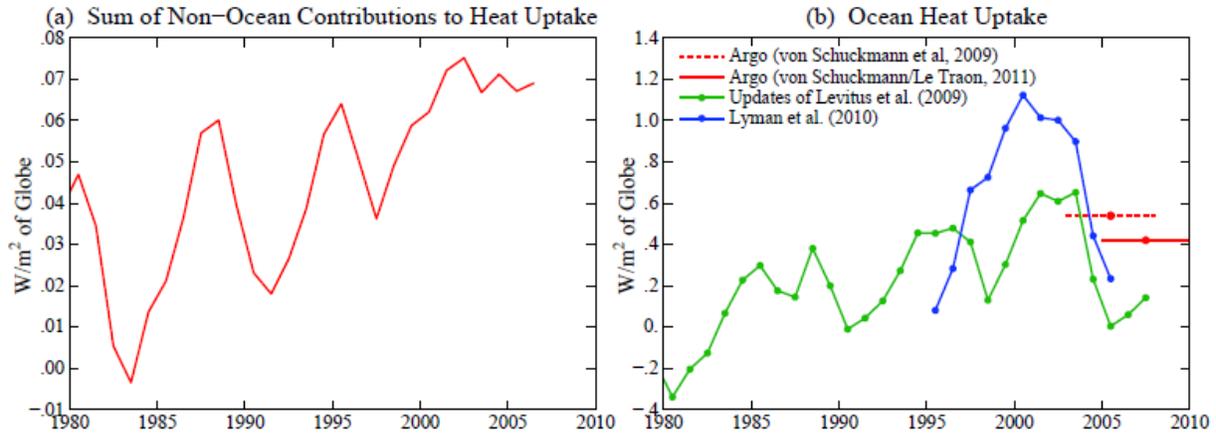

**Fig. 9.** (a) Sum of non-ocean contributions to planetary energy imbalance from Fig. 8, (b) Six-year trends of ocean heat uptake estimated by Levitus et al. (2009) and Lyman et al. (2010) for upper 700 m of the ocean, and estimates based on Argo float data for the upper 2000 m for 2003-2008 and 2005-2010.

### 9.2. Ocean term in planetary energy imbalance

Because of the ocean's huge heat capacity, temperature change must be measured very precisely to determine the ocean's contribution to planetary energy imbalance. Adequate precision is difficult to attain because of spatial and temporal sparseness of data, regional and seasonal biases in observations, and changing proportions of data from various instrument types with different biases and inaccuracies (Harrison and Carson, 2007; Domingues et al., 2008; Lyman and Johnson, 2008; Roemmich and Gilson 2009; Purkey and Johnson, 2010).

It has been possible to identify and adjust for some instrumental biases (e.g., Wijffels et al., 2008). These ameliorations have been shown to reduce what otherwise seemed to be unrealistically large decadal variations of ocean heat content (Dominigues et al., 2008).

Limitations in the spatial sampling and quality of historical ocean data led to deployment in the past decade of the international array of Argo floats capable of measurements to 2000 m (Roemmich and Gilson, 2009). Even this well-planned program had early instrumental problems causing data biases (Willis et al., 2007), but it was possible to identify and eliminate problematic data. Lyman and Johnson (2008) show that by about 2004 the Argo floats had sufficient space-time sampling to yield an accurate measure of heat content change in the upper ocean.

Graphs of ocean heat content usually show cumulative change. The derivative of this curve, the annual change of heat content, is more useful for our purposes, even though it is inherently "noisy". The rate of ocean heat uptake determines the planetary energy imbalance, which is the most fundamental single measure of the state of Earth's climate. The planetary energy imbalance is the drive for future climate change and it is simply related to climate forcings, being the portion of the net climate forcing that the planet has not yet responded to.

The noisiness of the annual energy imbalance is reduced by appropriate smoothing over several years. Von Schuckmann and Le Traon (2011) calculate a weighted linear trend for the 6-year period of most complete data, 2005-2010, the weight accounting for modest improvement in spatial coverage of observations during the 6-year period. They obtain a heat content trend of $0.55 \pm 0.1$ W/m$^2$ with analysis restricted to depths 10-1500 m and latitudes 60N-60S, equivalent

The uncertainty (standard error) for the von Schuckmann and Le Traon (2011) analyses does not include possible remaining systematic biases in the Argo observing system such as uncorrected drift of sensor calibration or pressure errors. Such biases caused significant errors in prior analyses. Estimated total uncertainty including unknown biases is necessarily subjective, but it is included in our summary below of all contributions to the planetary energy imbalance.

We emphasize the era of Argo data because of its potential for accurate analysis. For consistency with the von Schuckmann and Le Traon (2011) analysis we smooth other annual data with a 6-year moving linear trend. The 6-year smoothing is a compromise between minimizing the error and allowing temporal change due to events such as the Pinatubo volcano and the solar cycle to remain apparent in the record.

Fig. 9b includes heat uptake in the upper 700 m of the ocean based on analyses of Lyman et al. (2010) and Levitus et al. (2009). The 1993-2008 period is of special interest, because satellite altimetry for that period allows accurate measurement of sea level change.

Lyman et al. (2010) estimate average 1993-2008 heat gain in the upper 700 m of the ocean as $0.64 \pm 0.11$ W/m$^2$, where the uncertainty range is the 90 percent confidence interval. The error analysis of Lyman et al. (2010) includes uncertainty due to mapping choice, instrument (XBT) bias correction, quality control choice, sampling error, and climatology choice. Lyman and Johnson (2008) and Lyman et al. (2010) describe reasons for their analysis choices, most significantly the weighted averaging method for data sparse regions.

Levitus and colleagues (Levitus et al., 2000, 2005, 2009) maintain a widely used ocean data set. Lyman and Johnson (2008) suggest that the Levitus et al. objective analysis combined with simple volumetric integration in analyzing ocean heat uptake allows temperature anomalies to relax toward zero in data sparse regions and thus tends to underestimate ocean heat uptake. However, the oceanographic community has not reached consensus on a best analysis of existing data, so we compare our calculations with both Levitus et al. (2009) and Lyman et al. (2010).

Lyman et al. (2010) and Levitus et al. (2009) find smaller heat gain in the upper 700 m in the Argo era than that found in the upper 2000 m by von Schuckmann and Le Traon (2011), as expected[2]. Although the accuracy of ocean heat uptake in the pre-Argo era is inherently limited, it is clear that heat uptake in the Argo era is smaller than it was during the 5-10 years preceding full Argo deployment, as discussed by Trenberth (2009, 2010) and Trenberth and Fasullo (2010).

Heat uptake at ocean depths below those sampled by Argo is small, but not negligible. Purkey and Johnson (2010) find the abyssal ocean (below 4000 m) gaining heat at rate $0.027 \pm 0.009$ W/m$^2$ (average for entire globe) in the past three decades. Purkey and Johnson (2010) show that most of the global ocean heat gain between 2000 m and 4000 m occurs in the Southern Ocean south of the Sub-Antarctic Front. They estimate the rate of heat gain in the deep Southern Ocean (depths 1000-4000 m) during the past three decades[3] to be $0.068 \pm 0.062$ W/m$^2$. The uncertainties given by Purkey and Johnson (2010) for the abyssal ocean and Southern Ocean heat uptake are the uncertainties for 95 percent confidence.

---

[2] von Schuckmann and Le Traon (2011) find heat gain 0.45, 0.55 and 0.60 W/m$^2$ for ocean depths 0-700 m, 10-1500 m, and 0-2000 m, respectively, based on 2005-2010 trends. Multiply by 0.7 for global imbalance.
[3] The data span 1981-2010, but the mean time was 1992 for the first sections and 2005 for the latter sections, so the indicated flux may best be thought of as the mean for the interval 1992-2005.

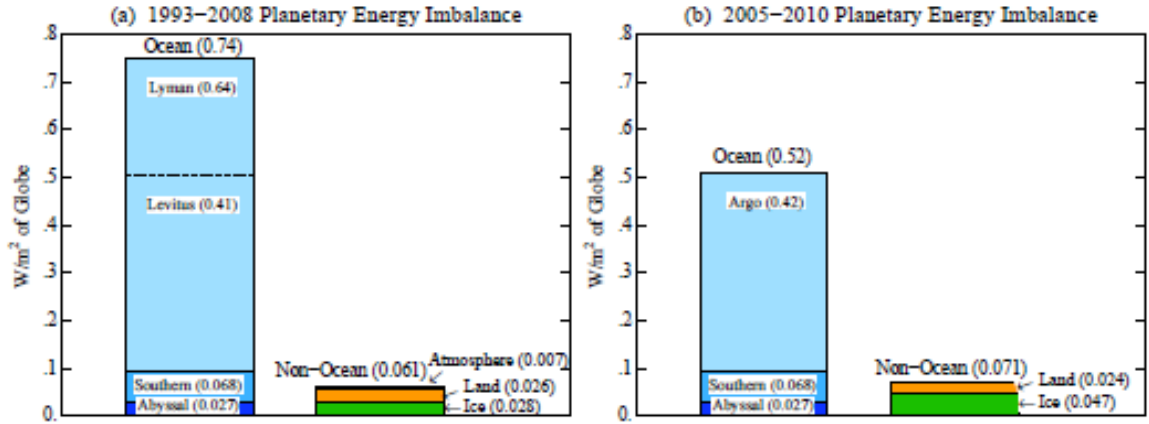

**Fig.10.** (a) Estimated contributions to planetary energy imbalance in 1993-2008, and (b) in 2005-2010. Except for heat gain in the abyssal ocean and Southern Ocean, ocean heat change beneath the upper ocean (top 700 m for period 1993-2008, top 2000 m in period 2005-2010) is assumed to be small and is not included. Data sources are the same as for Figs. 8 and 9 with discussion and references given in the text.

### 9.3. Summary of contributions to planetary energy imbalance

Knowledge of Earth's energy imbalance becomes increasingly murky as the period extends further into the past. Our choice for starting dates for summary comparisons in Fig. 10 is (a) 1993 for the longer period, because sea level began to be measured from satellites then, and (b) 2005 for the shorter period, because Argo floats had achieved nearly full spatial coverage.

Observed planetary energy imbalance includes upper ocean heat uptake plus three small terms. The first term is the sum of non-ocean terms (Fig. 9a). The second term, heat gain in the abyssal ocean (below 4000 m), is estimated to be $0.027 \pm 0.009$ W/m$^2$ by Purkey and Johnson (2010), based on observations in the past three decades. Deep ocean heat change occurs on long time scales and is expected to increase (Wunsch et al., 2007). Because global surface temperature increased almost linearly over the past three decades (Hansen et al., 2010) and deep ocean warming is driven by surface warming, we take this rate of abyssal ocean heat uptake as constant during 1980-present. The third term is heat gain in the ocean layer between 2000 and 4000 m for which we use the estimate $0.068 \pm 0.061$ W/m$^2$ of Purkey and Johnson (2010).

Upper ocean heat storage dominates the planetary energy imbalance during 1993-2008. Ocean heat change below 700 m depth in Fig. 10 is only for the Southern and abyssal oceans, but those should be the largest supplements to upper ocean heat storage (Leuliette and Miller, 2009). Levitus et al. (2009) depth profiles of ocean heat gain suggest that 15-20 percent of ocean heat uptake occurs below 700 m, which would be mostly accounted for by the estimates for the Southern and abyssal oceans. Uncertainty in total ocean heat storage during 1993-2008 is dominated by the discrepancy at 0-700 m between Levitus et al. (2009) and Lyman et al. (2010).

The Lyman et al. (2010) upper ocean heat storage of $0.64 \pm 0.11$ W/m$^2$ for 1993-2008 yields planetary energy imbalance 0.80 W/m$^2$. The smaller upper ocean heat gain of Levitus et al. (2009), 0.41 W/m$^2$, yields planetary energy imbalance 0.57 W/m$^2$.

The more recent period, 2005-2010, has smaller upper ocean heat gain, 0.38 W/m$^2$ for depths 10-1500 m (von Schuckmann and Le Traon, 2011) averaged over the entire planetary surface and 0.42 W/m$^2$ for depths 0-2000 m. The total planetary imbalance in 2005-2010 is 0.59 W/m$^2$. Non-ocean terms contribute 13 percent of the total heat gain in this period, exceeding the contribution in the longer period in part because of the increasing rate of ice melt.

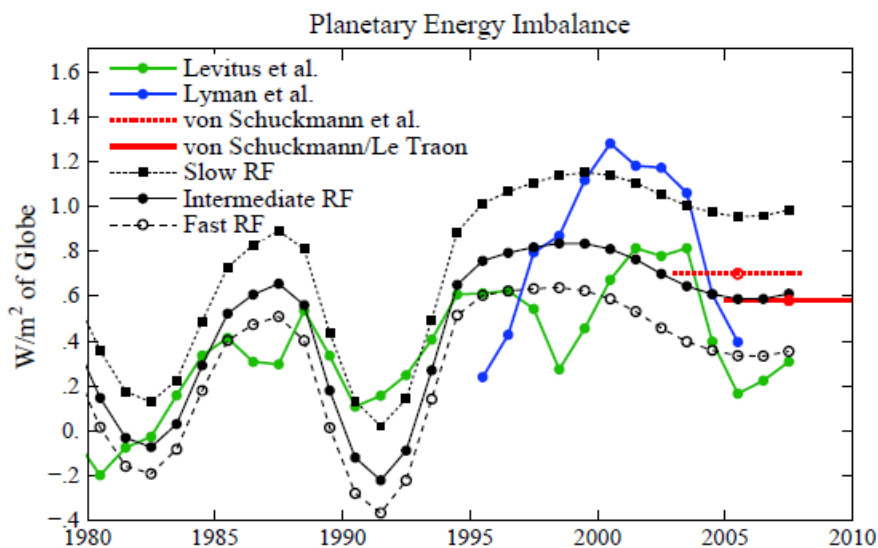

**Fig. 11.** Observed and calculated planetary energy imbalance, smoothed with moving 6-year trend. Non-ocean terms of Fig. 9a and small contributions of the abyssal ocean and deep Southern Ocean, as discussed in the text, are added to the upper ocean heat content analyses of Levitus et al. (2009), Lyman et al. (2010), von Schuckmann et al. (2009), and von Schuckmann and Le Traon (2011). Results for slow, intermediate, and fast climate response functions each fit the observed temperature record (Fig. 7).

Estimates of standard error of the observed planetary energy imbalance are necessarily partly subjective because the error is dominated by uncertainty in ocean heat gain, including imperfect instrument calibrations and the possibility of unrecognized biases. The von Schuckmann and Le Traon (2011) error estimate for the upper ocean (0.1 W/m$^2$) is 0.07 W/m$^2$ for the globe, excluding possible remaining systematic biases in the Argo observing system. Non-ocean terms (Fig. 8) contribute little to the total error because the terms are small and well defined. The error contribution from estimated heat gain in the deep Southern and abyssal oceans is also small, because the values estimated by Purkey and Johnson (2010) for these terms, 0.062 and 0.009 W/m$^2$, respectively, are their 95 percent (2-$\sigma$) confidence limits.

Our estimated planetary energy imbalance is $0.80 \pm 0.20$ W/m$^2$ for 1993-2008 and $0.59 \pm 0.15$ W/m$^2$ for 2005-2010, with estimated 1-$\sigma$ standard error. Our estimate for 1993-2008 uses the Lyman et al. (2010) ocean heat gain rather than Levitus et al. (2009) for the reason discussed in section 10. The estimated error in 2005-2010 is almost as large as that in 1993-2008 because of the brevity of the period. Sampling error in the Argo era will decline as the Argo record lengthens (von Schuckmann and Le Traon, 2011), but systematic biases may remain.

## 10. Modeled versus observed planetary energy imbalance

Fig. 11 compares observed and simulated planetary energy imbalances. Observations, smoothed with moving 6-year trends, are from Levitus et al. (2009), Lyman et al. (2010), von Schuckmann et al. (2009) and von Schuckmann and Le Traon (2011). The three small energy balance terms described above are added to the observed upper ocean heat uptake. Simulations are from Fig. 7, but smoothed with a moving 6-year trend to match smoothing of observations.

Argo era observed planetary energy imbalances are 0.70 W/m$^2$ in 2003-2008 and 0.59 W/m$^2$ in 2005-2010. Slow, intermediate, and fast response functions yield planetary energy imbalances 0.95, 0.59 and 0.34 W/m$^2$ in 2003-2008 and 0.98, 0.61 and 0.35 W/m$^2$ in 2005-2010.

Observed planetary energy imbalance in 1993-2008 is 0.80 W/m$^2$, assuming the Lyman et al. (2010) upper ocean heat storage, but only 0.59 W/m$^2$ with the Levitus et al. (2009) analysis. The calculated planetary energy imbalance for 1993-2008 is 1.06, 0.74 and 0.53 W/m$^2$ for the slow, intermediate and fast climate response functions, respectively.

We conclude that the slow climate response function is inconsistent with the observed planetary energy imbalance. This is an important conclusion because it implies that many climate models have been using an unrealistically large net climate forcing and human-made atmospheric aerosols probably cause a greater negative forcing than commonly assumed.

The intermediate response function yields planetary energy imbalance in close agreement with Argo-era observations. The intermediate response function also agrees with the planetary energy imbalance for 1993-2008, if we accept the Lyman et al. (2010) estimate for upper ocean heat uptake. Given that (1) Lyman et al. (2010) data is in much better agreement with the Argo-era analyses of von Schuckmann et al., and (2) a single response function must fit both the Argo-era and pre-Argo-era data, these results support the contention that the Levitus et al. analysis understates ocean heat uptake in data sparse regions. However, note that the conclusion that the slow response function is incompatible with observed planetary energy imbalance does not require resolving the difference between the Lyman et al. and Levitus et al. analyses.

Our principal conclusions, that the slow response function is unrealistically slow, and thus the corresponding net human-made climate forcing is unrealistically large, are supported by implications of the slow response function for ocean mixing. The slow response model requires a large net climate forcing (~2.1 W/m$^2$ in 2010) to achieve global surface warming consistent with observations, but that large forcing necessarily results in a large amount of heat being mixed into the deep ocean. Indeed, GISS modelE-R achieves realistic surface warming (Hansen et al., 2007b), but heat uptake by the deep ocean exceeds observations. Quantitative studies will be reported by others (A. Romanou and J. Marshall, private communication) confirming that GISS modelE-R has excessive deep ocean uptake of heat and passive tracers such as CFCs.

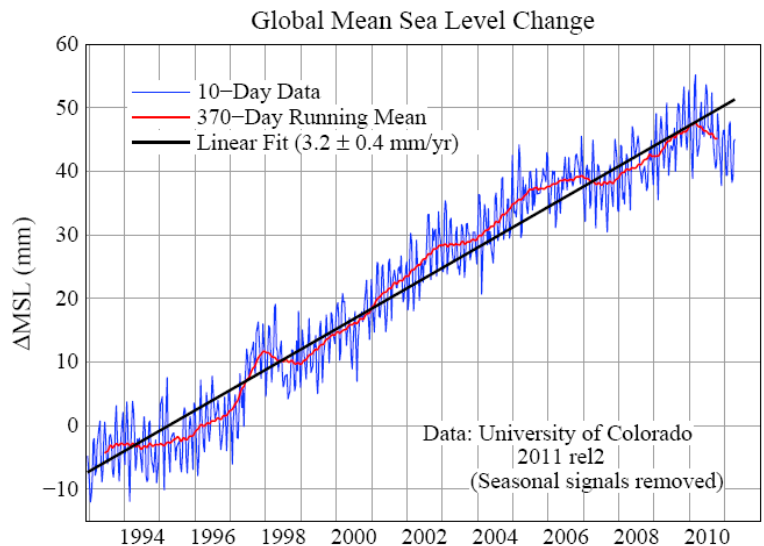

**Fig. 12.** Sea level change based on satellite altimeter measurements calibrated with tide-gauge measurements (Nerem et al., 2006; data updates at http://sealevel.colorado.edu/).

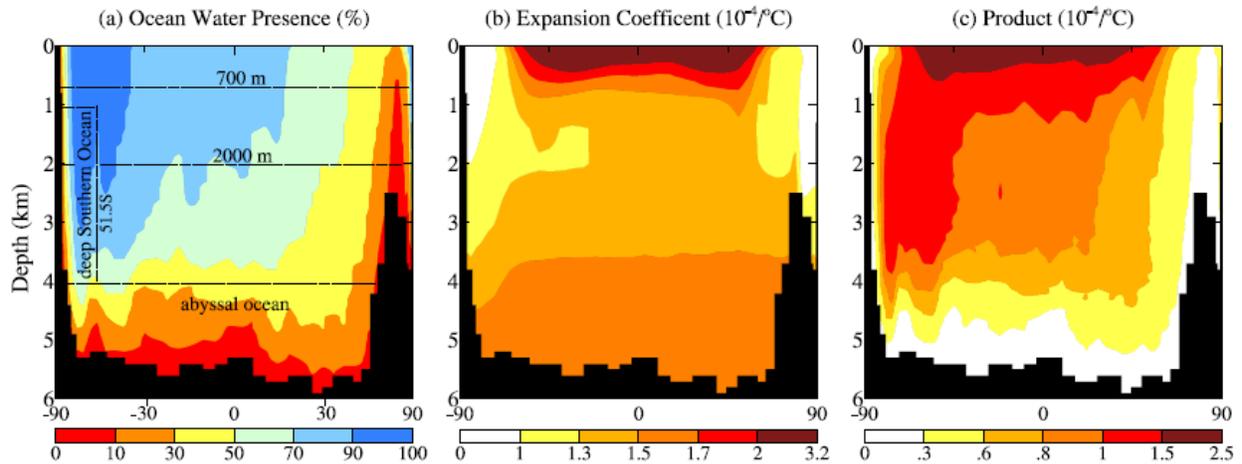

**Fig. 13.** (a) Percent of latitude-depth space occupied by water, (b) thermal expansion coefficient of water in today's ocean, (c) product of the quantities in (a) and (b). Equal intervals of the latitude scale have equal surface area. Calculations are area-weighted. Latitude ranges 90-60S, 60-30S, 30S-0, 0-30N, 30-60N, 60-90N contain 5, 26, 28, 27, 12, 2 percent of the ocean mass, and ocean surface in these latitude belts cover 4, 17, 19, 18, 9, 3 percent of the global surface area, respectively.

## 11. Is there closure with observed sea level change?

Munk (2002, 2003) drew attention to the fact that melting ice and thermal expansion of the ocean did not seem to be sufficient to account for observed sea level rise. This issue now can be reexamined with the help of Argo data and improving data on the rate of ice melt.

Fig. 12 shows sea level change in the period of global satellite observations (Nerem et al., 2006). Sea level increased at an average rate $3.2 \pm 0.4$ mm/year during 1993-2010. In the six year period of the most accurate Argo data, 2005-2010, sea level increased $2.0 \pm 0.5$ mm/year. The slower recent rate of sea level rise may be due in part to the strong La Nina in 2010.

Fig. 13 shows the potential of different volumes of the ocean to cause sea level rise via thermal expansion. The horizontal axis is proportional to cosine of latitude, so that equal increments have equal surface area on the planet. Fig. 13b shows that movement of heat from the tropical-subtropical upper ocean to greater depths, or especially to higher latitudes, by itself causes global sea level fall. The quantity in Fig. 13c must be multiplied by temperature change to find the contribution to ocean thermal expansion. Observed temperature change is largest in the upper few hundred meters of the ocean, which thus causes most of the sea level rise due to thermal expansion. Observed warming of the deep Southern Ocean and the abyssal ocean contributes a small amount to sea level rise (Purkey and Johnson, 2010).

Ocean temperature change in the upper 1500 m during 2005-2010 caused thermal expansion of $0.69 \pm 0.14$ mm/year (von Schuckmann and Le Traon, 2011). Warming of the deep (1000-4000 m) Southern Ocean and the abyssal ocean during the past three decades contributed at rates, respectively, $0.073 \pm 0.067$ and $0.053 \pm 0.017$ mm/year (Purkey and Johnson, 2010)[4]. Because global surface temperature increased almost linearly in recent decades (Hansen et al., 2010) and deep ocean warming is driven by surface warming, we take this mean rate of deep ocean warming as our estimate for these small terms. Thus thermal expansion in the Argo period contributes about 0.8 mm/year to sea level rise.

---

[4] This sea level rise due to Southern Ocean thermal expansion differs slightly from the published value as S. Purkey kindly recomputed this term (private comm., 2011) to eliminate overlap with Argo data.

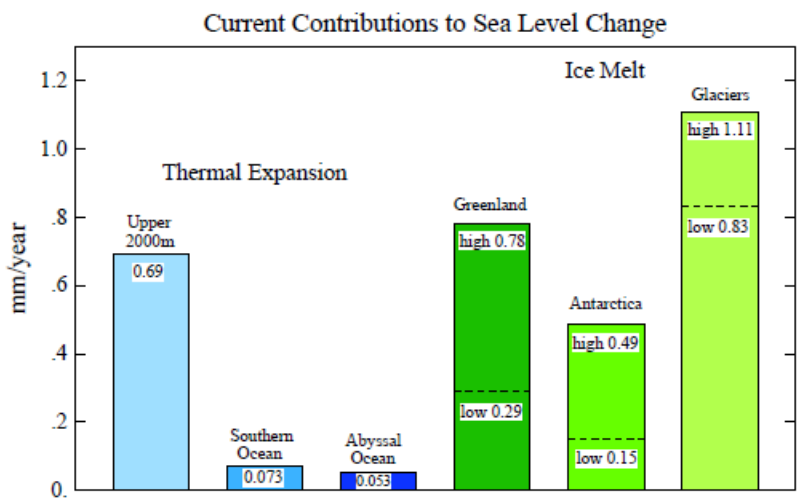

**Fig.14.** Estimated contributions to sea level change in 2005-2010. Upper ocean thermal expansion is from von Schuckmann and Le Traon (2011). High and low estimates for ice melt are discussed in the text.

Satellite measurement of Earth's changing gravity field should eventually allow accurate quantification of the principal contributions of ice melt to sea level rise. But at present there is a range of estimates due in part to the difficulty of disentangling ice mass loss from crustal isostatic adjustment (Bromwich and Nicolas, 2010; Sorensen and Forsberg, 2010; Wu et al., 2010). The 'high' estimates in Fig. 14 for Greenland and Antarctica, respectively, 281 and 176 Gt/year (360 Gt = 1 mm sea level), are from Velicogna (2009). A recent analysis (Rignot et al., 2011), comparing surface mass budget studies and the gravity method, supports the high estimates of Velicogna (2009). The low estimate for Greenland, 104 Gt/year, is from Wu et al. (2010). The low estimate for Antarctica, 55 Gt/year is the low end of the range -105 ± 50 Gt/year of S. Luthcke et al. (private communication, 2011). The high value for glaciers and small ice caps (400 Gt/year) is the estimate of Meier et al. (2007), while the low value (300 Gt/year) is the lower limit estimated by Meier et al. (2007).

Groundwater mining, reservoir filling, and other terrestrial processes also affect sea level. However, Milly et al. (2010) estimate that groundwater mining has added about 0.25 mm/year to sea level, while water storage has decreased sea level a similar amount, with at most a small net effect from such terrestrial processes. Thus ice melt and thermal expansion of sea water are the two significant factors that must account for sea level change.

The high value for total ice melt (857 Gt/year) yields an estimated rate of sea level rise of 0.80 (thermal expansion) + 2.38 (ice melt) = 3.18 mm/year. The low value for ice melt (459 Gt/year) yields 0.80 + 1.27 = 2.07 mm/year.

We conclude that ice melt plus thermal expansion are sufficient to account for observed sea level rise. Indeed, the issue now seems to be more the contrary of Munk's: why, during the years with data from both the gravity satellite and ARGO, is observed sea level rise so small?

Earth's energy imbalance provides information that is relevant to this question, because the planetary energy imbalance is the energy source for both ocean thermal expansion and melting of ice. Thus we must first examine the changing planetary energy imbalance, and then we will return to discussion of sea level rise in section 13.5.

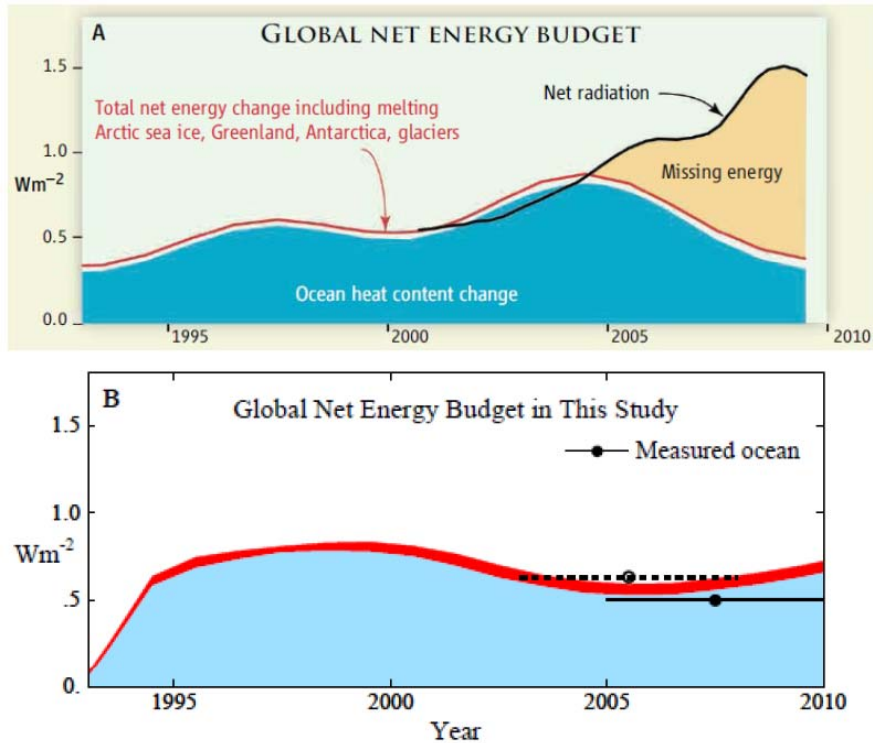

**Fig. 15.** Schematic diagram of Earth's energy imbalance (above) from Trenberth and Fasullo (2010) and (below) as calculated in this study. Our calculated imbalance, the top edge of the red area, is the moving 6-year trend of the calculated planetary energy imbalance for the intermediate response function. The red area is energy uptake by melting ice and warming ground and air. Measured ocean is 6-year trends of Argo analyses (see text) plus two small deep ocean terms from Purkey and Johnson (2010). Although the calculated decline of the energy imbalance is modest, it is significant because its origin in the solar irradiance minimum and Pinatubo volcanic aerosol rebound effect is insensitive to uncertainties in ocean mixing and climate sensitivity.

## 12. Why did planetary energy imbalance decline during the past decade?

The observed rate of ocean heat uptake since 2003 is less than in the preceding 10 years. Indeed, early reports suggested ocean cooling after 2003 (Lyman et al., 2006). That apparent cooling was a result of instrumental artifacts, but even after corrections the rate of heat uptake was smaller than in the prior decade (Willis et al., 2007). Observational error makes it difficult to measure heat uptake on short time scales, especially pre-Argo, but the slowdown in heat uptake since 2003 seems to be robust (Levitus et al., 2009; Lyman et al., 2010).

The slowdown of ocean heat uptake, together with satellite radiation budget observations, led to a perception that Earth's energy budget is not closed (Trenberth, 2009; Trenberth and Fasullo, 2010), as summarized in Fig. 15A. However, our calculated energy imbalance is consistent with observations (Fig. 15B), implying that there is no missing energy in recent years.

Note that, unlike Fig. 15B, real-world planetary energy imbalance includes unpredictable chaotic variability. A climate model with realistic interannual variability yields unforced interannual variability of global mean energy balance of 0.2-0.3 W/m$^2$ (Fig. 1, Hansen et al., 2005a). This 'noise' is eliminated in our calculations by the straight line representation of the climate response function (Fig. 5), but we must bear in mind unforced variability when interpreting observations. The unforced variability does not reduce the importance of the mean energy imbalance as a determinant of future climate.

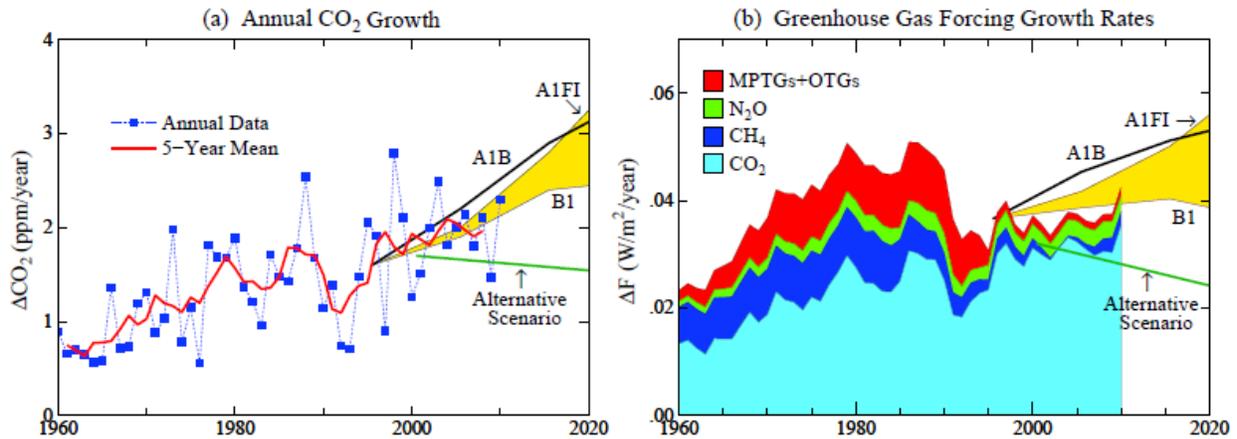

**Fig.16.** Annual growth of (a) atmospheric $CO_2$ and (b) climate forcing by long-lived greenhouse gases (GHGs). The GHG forcing is the 5-year running mean, except the final two points, for 2009 and 2010, are 3-year and 1-year means. For explanations see Hansen and Sato (2004).

In this section we examine why the calculated energy imbalance declined during the past decade. In section 13 we discuss factors that may account for the difference between expectations in Fig. 15A and the observed planetary energy imbalance.

Our calculated planetary energy imbalance is a function of only the climate forcings (Fig. 1) and the climate response function (Fig. 5). Relevant characteristics of the climate response function are the rapid initial response, about 40 percent within five years, and then the long slow 'recalcitrant' response, to use the adjective proposed by Held et al. (2010). The rapid response implies that even moderate ongoing changes of the climate forcings can have a noticeable effect, despite the fact that the climate system is still in a mode of trying to come to equilibrium with forcing changes that occurred over the past century.

## 12.1. Greenhouse gas climate forcing

Greenhouse gas (GHG) climate forcing would not seem to be a candidate to explain the recent dip in the planet's energy imbalance, because GHG forcing has increased monotonically. However, the growth rate of GHG forcing has experienced a relevant important change.

$CO_2$ is the main cause of increasing GHG forcing. Average $CO_2$ growth increased from 1 ppm (part per million) per year in the late 1960s to 2 ppm/year today (Fig. 16a). Contrary to a common misperception, $CO_2$ is not increasing faster than IPCC projections. Human-made $CO_2$ emissions are increasing just above the range of IPCC scenarios (Rahmsdorf et al., 2007), but the $CO_2$ increase appearing in the atmosphere, the 'airborne fraction' of emissions, has continued to average only about 55 percent (Supplementary Material, Hansen et al., 2008), despite concerns that the terrestrial and oceanic sinks for $CO_2$ are becoming less efficient (IPCC, 2007).

The annual increase of GHG climate forcing reached 0.05 W/m$^2$ in the late 1970s (Fig. 16b) but declined around 1990 as the growth of CFCs and $CH_4$ decreased (Hansen and Sato, 2004). MPTGs and OTGs in Fig. 16 are 'Montreal Protocol Trace Gases' and 'Other Trace Gases' delineated by Hansen and Sato (2004). Forcing in Fig. 16 is the commonly used 'adjusted' forcing (Ramaswamy et al., 2001; Hansen et al., 2005b). If this forcing is modified to also incorporate the varying 'efficacy' of each forcing, the only noticeable change is an increase of the $CH_4$ forcing by about 40 percent. Efficacy accounts for the fact that a $CH_4$ increase causes tropospheric $O_3$ and stratospheric $H_2O$ to increase. The choice of forcing definition has little effect on our present considerations because the recent growth of $CH_4$ has been slow. The definition has no effect on the comparison between the forcing for measured gas changes and the forcing for IPCC scenarios, because we use the same definition of forcing in all cases.

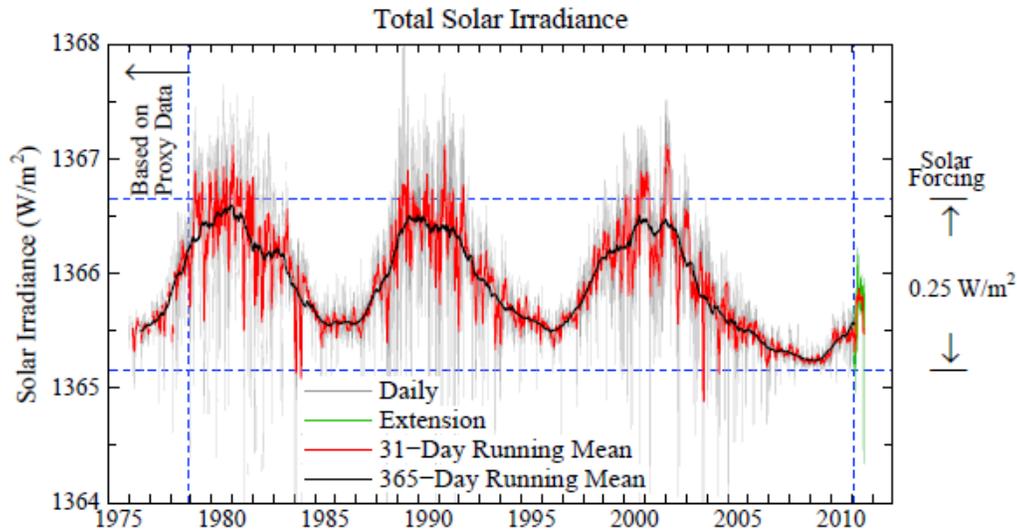

**Fig. 17.** Solar irradiance from composite of several satellite-measured time series. Data through 2 February 2012 is from Frohlich and Lean (1998 and <u>Physikalisch Meteorologisches Observatorium Davos, World Radiation Center</u>). Update in 2011 (through 24 August) is from <u>University of Colorado Solar Radiation & Climate Experiment</u> normalized to match means over the final 12 months of the Frohlich and Lean data.

### 12.2. Solar irradiance forcing

Solar irradiance has been measured from satellites since 1979. The continuous record in Fig. 17 required stitching together some barely overlapping satellite records (Frohlich and Lean, 1998). The longevity of the recent protracted solar minimum, at least two years longer than prior minima of the satellite era, makes that solar minimum potentially a potent force for cooling.

The amplitude of solar irradiance variability, measured perpendicular to the sun-Earth direction, is about 1.5 $W/m^2$ (left scale of Fig. 17), but because Earth absorbs only 240 $W/m^2$, averaged over the surface of the planet, the full amplitude of the solar forcing is only about 0.25 $W/m^2$. This is small compared to the human-made GHG forcing. But for the purpose of judging the effectiveness of solar variability on near-term climate change it is more appropriate to compare solar forcing with Earth's current energy imbalance, which is 0.6-0.7 $W/m^2$. It is thus apparent that the solar forcing is not negligible.

Solar forcing might be magnified by indirect effects. Solar variability at ultraviolet wavelengths alters stratospheric $O_3$, but any amplification of surface temperature response is at most about 10 percent (Shindell et al., 2001). Svensmark et al. (2009) suggest that solar activity may modulate terrestrial cloud cover. Studies by Calogovic et al. (2010) and Kumala et al. (2010) are not supportive, cloud chamber experiments of Kirkby et al. (2011) find a small effect of cosmic rays on aerosol nucleation, which conceivably could provide a mechanism for solar variability to modify clouds. Empirical correlation of ocean surface temperature with the solar cycle has been found with amplitude a few hundredths of a degree Celsius, consistent with solar forcing without any indirect amplification (White et al., 1997, 1998). Tung et al. (2008) argue that observed global temperature change in recent decades reveals a response in phase with solar irradiance change, with amplification up to a factor of two greater than expected from the direct solar forcing. However, the Tung et al. (2008) analysis does not fully remove the effect of volcanic eruptions that occurred approximately in phase with the solar cycle, so their inferred amplification is an upper limit on what is possible.

We use the measured solar variability (Fig. 17) to define the solar forcing for calculations without any amplification for indirect effects. However, we bear in mind that there remains a possibility that moderate amplification of the direct solar forcing exists.

### 12.3. Stratospheric aerosol forcing

Large volcanic eruptions can inject dust and sulfur dioxide gas into the stratosphere. Within months the $SO_2$ oxidizes, forming sulfuric acid aerosols that remain in the stratosphere for up to a few years (Robock, 2000). The aerosols reflect sunlight, causing a negative (cooling) climate forcing. Stratospheric aerosols were precisely monitored from satellites in much of the past three decades by viewing the sun through Earth's atmosphere (McCormick et al., 1995).

We use the stratospheric aerosol history compiled by Sato et al. (1993) and updates estimated with the aid of information provided by L. Thomason (private communication, 2008) and Haywood et al. (2010). Aerosol optical depth for tropical eruptions is assumed to increase linearly to a maximum 4 months after the eruption and then decay exponentially with 1-year e-folding time. High latitude eruptions reach a maximum in 3 months and decay with 3-month e-folding time. Peak global mean optical depth for tropical eruptions is taken as 0.01 for Ruang (Sept. 2002), 0.01 for Manam (Jan. 2005), 0.002 for Soufrier (May 2006), 0.0023 for Kasatochi (Aug. 2008), and 0.005 for Sarychev (June 2009). High latitude eruptions (Kasatochi and Sarychev) are assumed to cover only ¼ of the globe (30-90°N) The updated aerosol history is available at http://data.giss.nasa.gov/modelforce/strataer/ The effective forcing is taken as

$$F(\tau) = -23\tau, \qquad (4)$$

where $\tau$ is the global mean aerosol optical depth, as discussed by Hansen et al. (2005b).

The ability of radiation calculations to simulate the effect of stratospheric aerosols on reflected solar and emitted thermal spectra was tested (Fig. 11 of Hansen et al., 2005) using Earth's measured radiation balance (Wong et al., 2005) following the 1991 Pinatubo eruption. Good agreement was found with the observed temporal response of solar and thermal radiation, but the observed net radiation change was about 25 percent smaller than calculated. Because of uncertainties in measured radiation anomalies and aerosol properties, that difference is not large enough to define any change to the stratospheric aerosol forcing. It is noteworthy that reduction of volcanic aerosol forcing by that amount would bring simulated temperatures after historic eruptions into closer agreement with observations (Robock, 2000; Hansen et al., 1996), but given the absence of further evidence we employ the above equation for aerosol forcing.

An alternative stratospheric aerosol time series for recent decades has recently been suggested by Solomon et al. (2011). Their stratospheric aerosol optical depth is smaller than ours at times of some of the specific volcanic eruptions listed above, but otherwise generally larger than ours because of an inferred growing background stratospheric aerosol amount. We repeated the computations of section 12.4 using the aerosol optical depth from Fig. 2 of Solomon et al. (2011). As expected, because the change of aerosol optical depth is ~0.005, the effect of this alternative stratospheric aerosol history is small and does not noticeably alter the calculated decadal energy imbalance or the inferred tropospheric aerosol optical depth, the latter being much larger than the change of stratospheric aerosol optical depth.

**12.4. Simulated surface temperature and energy imbalance**

Fig. 18 examines the effect of each of the climate forcings on global mean temperature and planetary energy imbalance. All cases use the intermediate climate response function of Fig. 5, which yields the most realistic ocean heat storage.

Fig. 18a shows that the combination of all forcings yields a decline in the planetary energy imbalance over the past decade. Here we will determine which specific forcings are responsible for this decline of the energy imbalance.

Figs. 18b, 18c, and 18d show the effect of GHGs, tropospheric aerosols, and the combination of both of these forcings. These two principal forcings determine the trend of global temperature over the past century.

GHG plus tropospheric aerosols yield a flat planetary energy imbalance beginning a quarter century ago at a level of about 0.6 W/m$^2$, less than the maximum imbalance of about 1 W/m$^2$ due to all forcings. That flattening is a consequence of the fact that the growth rate of the GHG forcing stopped increasing about 1980 and then declined to a lower level about 1990 (Fig. 16b). The flat planetary energy balance due to these principle forcings allows small negative forcings in the past decade and the Pinatubo eruption 20 years ago to cause first a rise of the planet's energy imbalance and then a decline.

Fig. 18e shows the effect of volcanic aerosols. Volcanoes cause a negative planetary energy imbalance during the 1-2 years that the aerosols are present in the stratosphere, followed by a rebound to a positive planetary energy imbalance. This rebound is most clearly defined after the Pinatubo eruption, being noticeable for more than a decade, because of the absence of other volcanoes in that period.

The physical origin of the rebound is simple. Solar heating of Earth returns to its pre-volcano level as aerosols exit the stratosphere. However, thermal emission to space is reduced for a longer period because the ocean was cooled by the volcanic aerosols. In calculations via the response function, using equation (2), the volcanic aerosols introduce a dF/dt of one sign and within a few years a dF/dt of opposite sign. The integrated (cumulative) dF/dt due to the volcano is zero but the negative dF/dt occurred earlier, so its effect on temperature, defined by the climate response function, is greater. The effect of the temporal spacing between the negative and positive changes of F decreases as time advances subsequent to the eruption.

The vertical scale for the solar irradiance forcing and its induced planetary energy imbalance are magnified by a factor 10 in Fig. 18g. The imbalance is in phase with the irradiance, but temperature maxima and minima lag irradiance maxima and minima by 1-2 years. The lag is caused by the ocean's thermal inertia, but the magnitude of the lag incorporates, via the response function, the effect of both the ocean and continental responses to the forcing.

The reduction of planetary energy imbalance between 2000 and 2009 due to declining solar irradiance is about 0.14 W/m$^2$. If there is an indirect effect magnifying the solar forcing, the calculated effect on the planetary energy imbalance must be increased by that magnification factor. As discussed in section 12.2, empirical correlations of the solar cycle and global temperature show that any magnification cannot exceed a factor of two at most.

In summary, precipitous decline in the growth rate of GHG forcing about 25 years ago caused a decrease in the rate of growth of the total climate forcing and thus a flattening of the planetary energy imbalance over the past two decades. That flattening allows the small forcing due to the solar cycle minimum, a delayed bounceback effect from Pinatubo cooling, and recent small volcanoes to cause a decrease of the planetary energy imbalance over the past decade.

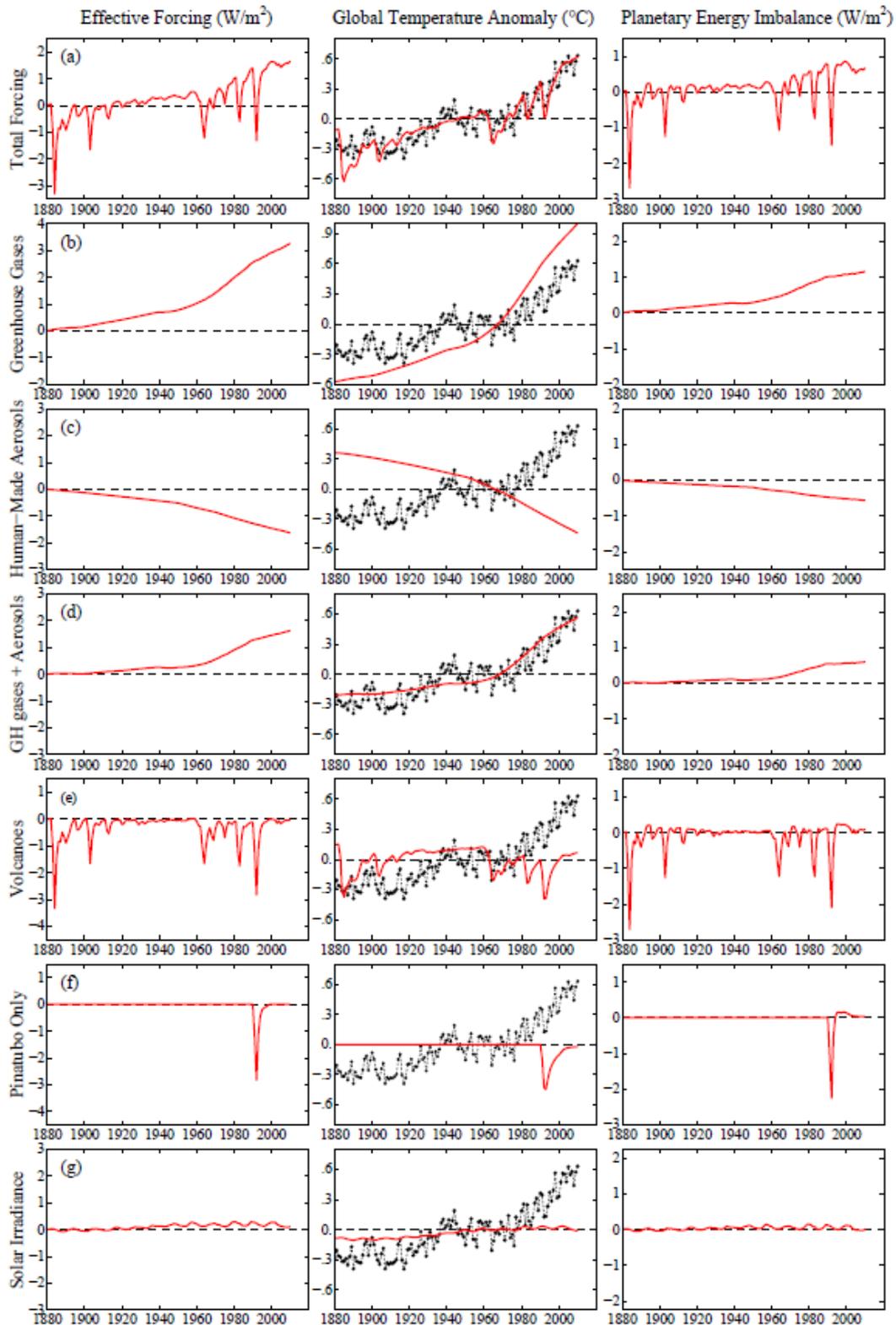

**Fig. 18.** Climate forcings and their contributions (red curves) to temperature change and planetary energy imbalance. Observed global temperature change is included in the middle column for comparison; base period is 1951-1980 (zero mean) for observations and model. Vertical scale is magnified by 10 for solar forcing and its contribution to planetary energy imbalance (g).

## 13. Discussion

Earth's energy imbalance, averaged over several years, is a fundamental characterization of the state of the climate. It determines how much additional global temperature change is 'in the pipeline'. It defines how much current climate forcings must be altered to stabilize climate.

Earth's average energy imbalance is expected to be only ~1 W/m$^2$ or less (IPCC, 2007; Hansen et al., 2005a). Therefore assessment of the imbalance requires measurement accuracy approaching 0.1 W/m$^2$. That target accuracy is just becoming conceivable, for data averaged over several years, with global distribution of Argo profiling floats. Measurements of Earth's energy imbalance will be invaluable for policy and scientific uses, if the observational system is maintained and enhanced. Implications of the data are discussed here.

### 13.1. Human-made climate forcing versus solar variability

Argo floats achieved good global distribution just in time for a valuable test of the effect of solar variability on Earth's energy imbalance and climate. The last half of the first decade of the 21$^{st}$ century witnessed the deepest most prolonged solar minimum in the period of accurate solar monitoring that began in the late 1970s (Fig. 17).

Earth's energy imbalance during the solar minimum tests the effect of solar variability on climate, including any amplifications that may exist, such as the effect of cosmic rays on clouds. The imbalance during solar minimum is the net effect of reduced solar irradiance and all other climate forcings, principally the net human-made climate forcing. Volcanic aerosols added a small negative forcing (Fig. 18e) that assisted the negative solar forcing.

Human-made forcing has been growing for more than a century and thus has partially expended itself, causing most of the 0.8°C global warming of the past century. However, because of the ocean's thermal inertia, the climate system has only partly responded to the human-made forcing. The portion of the human-made forcing that has not been responded to constitutes a continuing forcing with positive sign (incoming energy exceeds outgoing energy). During the past 5-6 years the deep solar minimum caused a negative forcing[5]. Precise measurement of the planetary energy imbalance allows us to determine whether the positive human-made forcing or negative natural forcing is larger.

A verdict is provided by the ocean heat uptake found by von Schuckmann and Le Traon (2011), 0.42 W/m$^2$ for 2005-2010, averaged over the planet[6]. Adding the small terms for heat uptake in the deeper ocean, warming of the ground and atmosphere, and melting of ice, the net planetary energy imbalance exceeded +0.5 W/m$^2$ during the solar minimum.

The strong positive energy imbalance during the solar minimum, and the consistency of the planet's energy imbalance with expectations based on estimated human-made climate forcing, together constitute a smoking gun, a fundamental verification that human-made climate forcing is the dominant forcing driving global climate change. Positive net forcing even during solar minimum assures that global warming will be continuing on decadal time scales.

---

[5] Solar irradiance is negative in 2005-2010 relevant to the mean during the period of measurements (Fig. 17). Recent solar forcing is positive relevant to 1880 (Fig. 18g) in the scenario used by Hansen et al. (2007), which is based on Lean (2000). The reality of the solar irradiance increase between 1900 and 1940, based on proxy indicators of solar activity, has been called into question by Lean et al. (2002). If the 1900-1940 solar irradiance increase is real, at least half of its effect would already have been expended in global warming, thus leaving a small negative contribution to planetary energy imbalance in 2005-2010 (Fig. 18g).

[6] von Schuckmann and Le Traon (2011) find 0.55 ± 0.1 W/m$^2$ for the top 1500 m of the ocean. Repeating their analysis for the top 2000 m yields 0.60 ± 0.1 W/m$^2$ (this paper), corresponding to 0.42 ± 0.07 W/m$^2$ globally.

**13.2. Climate response function**

Earth's climate response function, the fraction of global surface temperature response to a climate forcing as a function of time, is a fundamental characteristic of the climate system that needs to be accurately determined. Climate models indicate that the response function has a characteristic shape, achieving almost half of its equilibrium response quickly, within about a decade. The remainder of the response is exceedingly slow. We suggest, however, that this recalcitrance is exaggerated in many climate models.

GISS modelE-R, for example, achieves only 60 percent response in 100 years. At least several other climate models used in IPCC (2001, 2007) studies have comparably slow response. Diagnostic studies of the GISS ocean model show that it mixes too efficiently, which would cause its response function to be too slow. Therefore we tested alternative response functions that achieve 75 percent and 90 percent of their response after 100 years. In each case we let current human-made aerosol forcing have the magnitude that yields closest agreement with observed global warming over the past century.

The amount of energy pumped into the ocean by a positive (warming) forcing depends on the response function, because this energy source (the planetary energy imbalance) shuts down as the surface temperature approaches its equilibrium response. Thus precise measurement of the rate of ocean heat uptake can help discriminate among alternative response functions.

Ocean heat uptake during the Argo era agrees well with the intermediate response function (75 percent response in 100 years) and is inconsistent with either the slow or fast response functions. We conclude that actual climate response is not as recalcitrant as many models suggest. As the Argo record lengthens, knowledge of the real world response function can be refined. The depth distribution of warming will be especially useful for confirming the nature of the climate response function, characterizing its recalcitrant response, and aiding development of ocean models.

**13.3. Aerosol climate forcing**

Aerosol climate forcing is unmeasured. Aerosol uncertainty is the principal barrier to quantitative understanding of ongoing climate change. Until aerosol forcing is measured, its magnitude will continue to be crudely inferred, implicitly or explicitly, via observations of climate change and knowledge of climate sensitivity.

We explicitly allowed aerosol forcing and the climate response function to be free variables. We used observations of global temperature change and ocean heat uptake to define the aerosol forcing and response function that yield best agreement with observations.

We assumed that aerosol forcing as a function of time had the shape of the aerosol curve in Fig. 1. Aerosol forcing for 1880-1990, described by Hansen et al. (2007), is based on aerosol modeling (Koch, 2001) using aerosol emissions from fuel use statistics and including temporal changes in fossil fuel technologies (Novakov et al., 2003). Our extension post-1990 assumed that aerosol forcing was half as large and opposite in sign of the GHG forcing, as was the case in the prior decade. That crude assumption is consistent with moderate reduction of aerosol amount in developed countries and increasing aerosols in developing countries.

Our derived aerosol forcing in 2010 is $-1.6$ W/m$^2$. This inferred aerosol forcing does not exceed IPCC (2007) a priori estimated aerosol forcing including all indirect effects. More important, it is consistent with an insightful study of Earth's energy balance based on satellite measurements of reflected solar radiation and emitted heat radiation (Murphy et al., 2009), as discussed in section 13.6.

Our derived aerosol forcing does exceed aerosol forcings employed in most climate simulations carried out for IPCC (2001, 2007). For example, an ensemble of models from several groups (Fig. 9 of Stott and Forest, 2007) had aerosol forcings in the range 0.4 to 1.1 W/m$^2$. Our interpretation of why these models produced agreement with observed temperature change over the past century is that the ocean models have a slow response function, slower than the real world, mixing heat too efficiently into the deep ocean.

### 13.4. Implications for climate stabilization

Earth is out of energy balance by at least 0.5 W/m$^2$. If other forcings are unchanged, atmospheric $CO_2$ must be reduced 30 ppm, to a level approximately 360 ppm, to increase Earth's heat radiation to space by 0.5 W/m$^2$.

However, the measured energy imbalance was 0.59 W/m$^2$ in 2005-2010, during a deep solar minimum. We estimate the energy imbalance averaged over a solar cycle as ~ 0.75 W/m$^2$. In that case, $CO_2$ would need to be reduced to about 345 ppm to restore energy balance, if other factors are fixed.

Other factors are not fixed, but $CO_2$ is the dominant forcing of long-term climate. It should be practical to keep the net effect of other human-made climate forcings close to zero, provided $CO_2$ and global warming are limited. Potential reduction of human-made climate forcing by $CH_4$, CFCs and black soot can largely compensate for the increase in forcing that will occur as an expected decrease of human-made reflective aerosols occurs (Hansen et al., 2000; Ramanathan et al., 2001; Jacobson, 2001). However, if $CO_2$ and global warming are not limited, release of $CH_4$ via melting of tundra and methane hydrates may frustrate attempts to prevent growth of non-$CO_2$ forcings.

Thus a target $CO_2$ level of 350 ppm (Hansen et al., 2008) is an appropriate initial goal for climate stabilization. Refinements can be made later, as the world approaches that goal.

Exact restoration of planetary energy balance is not necessarily the optimum target to achieve long-term climate stabilization. Global warming already in place may have undesirable effects via slow feedbacks, for example on ice sheet stability and sea level. Such issues must be evaluated, for example via continued monitoring of ice sheet mass balance, as planetary energy balance is approached. A moderately negative planetary energy imbalance may be needed to stabilize sea level. Such refinements will become a practical issue only after GHGs have been reduced to a level that approximates planetary energy balance.

### 13.5. Implications for sea level

Sea level rise has been about 3 mm/year since satellite measurements began in the early 1990s (Fig. 12). Thermal expansion of ocean water and ice melt can account sea level rise of that magnitude. Assuming that the nearly constant rate of sea level rise in Fig. 12 is accurate, the near constancy is perhaps a consequence of an increasing rate of ice melt and decreasing thermal expansion. Annual thermal expansion of the ocean is expected to have been maximum in 1993 because ocean cooling due to the 1991 Pinatubo eruption would have peaked, leaving the gap between equilibrium and actual global temperature at a maximum. Although there is large interannual variability associated with the El Nino/La Nina cycle, thermal expansion should have been on a downward trend from 1993 to the present. Ice melt, in contrast, has probably been increasing over the period 1993-present (Fig. 8c).

Curiously, the rate of sea level rise seems to have slowed in the past six years (2005-2010) to about 2 mm/year (Fig. 12), despite the apparently increasing rate of mass loss from the Greenland and Antarctic ice sheets (Fig. 8c). The slower rate of sea level rise during 2005-2010 is probably due at least in part to the La Nina in 2009-2010, but the effect of the La Nina should be captured in the calculation of 0.80 mm/year thermal expansion based on Argo data. Thus the recent measured sea level rise favors the lower estimates of ice sheet melt.

The low (2 mm/year) rate of sea level rise is not likely to continue. Based on our inferred planetary energy imbalance, we conclude that the rate of sea level rise should accelerate during the next several years. Reasons for that conclusion are as follows.

First, the contribution of thermal expansion to sea level is likely to increase above recent rates. Solar minimum and a diminishing Pinatubo rebound effect both contributed to a declining rate of thermal expansion during the past several years. But the Pinatubo effect is now essentially spent and solar irradiance change should now work in the opposite sense.

Second, the rate of ice melt is likely to continue to accelerate. Planetary energy imbalance now is positive, substantial, and likely to increase as greenhouse gases and solar irradiance increase. Thus, despite year-to-year fluctuations, global temperature will increase this decade and there will be a substantial flux of energy into the ocean. Increasing ocean heat content provides energy for melting sea ice and ice shelves. Sea ice protects the ice sheets from heating and ice shelves mechanically buttress the ice sheets. It has been argued that loss of these protections of the surrounding ice may be the most important factor causing more rapid discharge from ice sheets to the ocean (Hansen, 2005, 2007).

**13.6. Implications for observations**

Earth's energy imbalance and its changes will determine the future of Earth's climate. It is thus imperative to measure Earth's energy imbalance and the factors that are changing it.

**13.6.1. Measuring Earth's energy imbalance**

The required measurement accuracy is ~0.1 W/m$^2$, in view of the fact that estimated current (2005-2010) energy imbalance is 0.59 W/m$^2$. The accuracy requirement refers to the energy imbalance averaged over several years. It is this average imbalance that drives future climate. Stabilization of climate requires the energy imbalance averaged over El Nino-La Nina variability and the solar cycle to be close to zero.

There are two candidate measurement approaches: (1) satellites measuring the sunlight reflected by Earth and heat radiation to space, (2) measurements of changes in the heat content of the ocean and the smaller heat reservoirs on Earth. Each approach has problems. There is merit in pursuing both methods, because confidence in the result will become high only when they agree or at least the reasons that they differ are understood.

The difficulty with the satellite approach becomes clear by considering first the suggestion of measuring Earth's reflected sunlight and emitted heat from a satellite at the Lagrange L1 point, which is a location between the sun and Earth at which the gravitational pulls from these bodies are equal and opposite. From this location the satellite would continually stare at the sunlit half of Earth.

The notion that a single satellite at this point could measure Earth's energy imbalance to 0.1 W/m$^2$ is prima facie preposterous. Earth emits and scatters radiation in all directions, i.e., into $4\pi$ steradians. How can measurement of radiation in a single direction provide a proxy for radiation in all directions? Climate change alters the angular distribution of scattered and emitted radiation. It is implausible that changes in the angular distribution of radiation could be modeled to the needed accuracy, and the objective is to measure the imbalance, not guess at it. There is also the difficulty of maintaining sensor calibrations to accuracy 0.1 W/m$^2$, i.e., 0.04 percent. That accuracy is beyond the state-of-the art, even for short periods, and that accuracy would need to be maintained for decades.  There are many useful measurements that could be made from a mission to the Lagrange L1 point, but Earth's radiation balance in not one of them.

These same problems, the changing angular distribution of the scattered and emitted radiation fields and maintaining extreme precision of sensors over long periods, must be faced by Earth-orbiting satellites.  Earth radiation budget satellites have progressed through several generations and improved considerably over the past half-century, and they provide valuable data, e.g., helping to define energy transport from low to high latitudes.  The angular distribution problem is treated via empirical angular distribution models, which are used to convert measurements of radiation in a given direction into radiative (energy) fluxes.

The precision achieved by the most advanced generation of radiation budget satellites is indicated by the planetary energy imbalance measured by the ongoing CERES (Clouds and the Earth's Radiant Energy System) instrument (Loeb et al., 2009), which finds a measured 5-year-mean imbalance of 6.5 W/m$^2$ (Loeb et al., 2009).  Because this result is implausible, instrumentation calibration factors were introduced to reduce the imbalance to the imbalance suggested by climate models, 0.85 W/m$^2$ (Loeb et al., 2009).

The problems being addressed with this tuning probably involve the high variability and changes of the angular distribution functions for outgoing radiation and the very limited sampling of the radiation field that is possible from an orbiting satellite, as well as, perhaps, detector calibration.  There can be no credible expectation that this tuning/calibration procedure can reduce the error by two orders of magnitude as required to measure changes of Earth's energy balance to an accuracy of 0.1 W/m$^2$.

These difficulties do not imply that attempts to extract the Earth's radiation imbalance from satellite measurements should not be continued and improved as much as possible.  The data are already useful for many purposes, and their value will only increase via continued comparisons with other data such as ocean heat uptake.

An alternative potentially accurate approach to measure Earth's energy imbalance is via changes in the ocean heat content, as has been argued for decades (Hansen et al., 1997) and as is now feasible with Argo data (Roemmich and Gilson, 2009; Von Schuckmann and Le Traon, 2011).  This approach also has sampling and instrument calibration problems, but it has a fundamental advantage: it is based on absolute measurements of ocean temperature.  As a result, the accuracy improves as the record length increases, and it is the average energy imbalance over years and decades that is of greatest interest.

The error estimated by von Schuckmann and Le Traon (2011) for ocean heat uptake in the upper 2000 m of the ocean, ± 0.1 W/m$^2$ for the ocean area or ± 0.07 W/m$^2$ for the planetary energy imbalance, does not include an estimate for any remaining systematic calibration errors that may exist.  At least some such errors are likely to exist, so continuing efforts to test the data and improve calibrations are needed.  The Argo program needs to be continued and expanded to achieve further improvement and minimization of error.

The Argo floats drift in location and have finite lifetime, so it is necessary to continually add about 800 floats per year to maintain the system.  Ocean south of 60S (includes 5.6 percent of the ocean area) and north of 60N (4.7 percent of ocean area) is not well sampled by Argo.  Floats capable of operating under sea ice need to be deployed in the polar oceans, and floats capable of extending the measurements into the deep and abyssal oceans need to be developed.  That more spatially complete data would help define the nature of the climate response function, characterize the climate system's recalcitrant response, and aid development of ocean models.

**13.6.2. Measuring the cause of Earth's energy imbalance**

We also must quantify the causes of changes of Earth's energy imbalance. The two dominant causes are changes of greenhouse gases, which are measured very precisely, and changes of atmospheric aerosols. It is remarkable and untenable that the second largest forcing that drives global climate change remains unmeasured. We refer to the direct and indirect effects of human-made aerosols.

We have inferred indirectly, from the planet's energy imbalance and global temperature change, that aerosols are probably causing a forcing of about −1.6 W/m$^2$ in 2010. Our estimated uncertainty, necessarily partly subjective, is ± 0.3 W/m$^2$, thus a range of aerosol forcing from −1.3 to −1.9 W/m$^2$.

Our conclusion can be compared with an insightful analysis of Murphy et al. (2009), which uses measurements of ocean heat content, greenhouse gases, volcanic aerosols, and correlations between surface temperature and satellite radiative flux measurements to infer a residual planetary energy flux that they presume to be caused by aerosol direct and indirect radiative forcing. Their result is an average aerosol forcing of −1.1 ± 0.4 W/m$^2$ for the period 1970-2000. For that period the aerosol forcing that we find (Fig. 1) is −1.2 ±0.3 W/m$^2$. The time dependence of the residual flux imbalance found by Murphy et al. (2009) is shown in their Fig. 4(c). It has an imbalance of about −1.5 W/m$^2$ in 2000 at the end of their analysis, which is consistent with our analysis.

These analyses tend to confirm that aerosol forcing is large and negative, but cannot tell us what aerosols are causing the forcing, how much of the forcing is due to indirect effects on clouds, and how the aerosol forcing is changing. Aerosol climate forcing is complex (Ramanathan et al., 2001; Ramaswamy et al., 2001), in part because there are many different aerosol compositions distributed inhomogeneously around the planet. Different compositions have different effects on solar radiation, and, via their effects on clouds, they have different effects on terrestrial thermal radiation.

Determination of the aerosol climate forcing requires measuring the aerosol physical properties and how those properties are changing. This is analogous to how climate forcing by greenhouse gases is determined. We cannot determine the greenhouse gas forcing by measuring the radiation within the atmosphere or from a satellite – there are too many factors that affect the radiation field, including climate changes, cloud changes, etc. Instead we measure the changes of $CO_2$, CFCs and other gases. We then compute, from basic physics, the climate forcing with an accuracy of better than 10 percent; even higher accuracy is possible, but not essential.

Existing satellite measurements provide an estimate of aerosol optical depth (Mishchenko and Geogdzhayev, 2007; Mishchenko et al., 2007b), but maps of this quantity show decreases and increases in various regions, some presumably dominated by reflecting sulfates, some by partially absorbing dust, some by weakly or heavily absorbing organic particles or black soot. These existing measurements do not determine aerosol climate forcing and how it is changing.

It has been shown, from theory, aircraft observations, and planetary studies (Mishchenko et al., 2007a) that as many as 10 parameters defining aerosol properties can be obtained using satellite measurements that fully characterize reflected solar radiation. Full characterization requires: (1) measurement of the linear polarization of the reflected radiation to an accuracy of the order of 0.1 percent, (2) measurement at several spectral bands spanning reflected solar radiation from the near-ultraviolet to the near-infrared, (3) measurement of the radiation over the full range of scattering angles available by scanning from horizon to horizon along the satellite ground track. Such measurements suffice to determine the direct aerosol climate forcing.

Polarization measurements of reflected sunlight also yield information about cloudtop droplets. However, determination of the aerosol indirect climate forcing requires precise measurement of small changes in clouds induced by aerosol changes. Cloud particles are one to two orders of magnitude larger than aerosols and thus clouds can be characterized best via measurement of thermal emission at wavelengths 5-50 µm (200 - 2000 cm$^{-1}$). The essential requirement is very high wavelength-to-wavelength precision, because such high precision data can yield atmospheric and cloud temperatures, cloud particle properties, and water vapor profile. The appropriate instrument for achieving such precision is a Michelson interferometer, because it records the signal from nearby wavelengths on the same detector.

The Glory mission (Mishchenko et al., 2007a), which was expected to begin operations this year, would have measured the aerosol direct forcing, as it carried an instrument capable of measuring polarization to an accuracy about 0.1 percent. However, launch failure caused loss of the satellite, which failed to achieve orbit. A replacement mission is being planned with launch expected in the 2015-2016 time frame. Such detailed composition-specific global aerosol measurements will be essential to interpret changing planetary energy balance. Presently the net effect of changing emissions in developing and developed countries is highly uncertain.

The large aerosol forcing derived in our present study implies that the aerosol indirect climate forcing exceeds the direct aerosol forcing, possibly by a large amount. There is no simple relationship between direct and indirect forcings, which each strongly dependent on aerosol composition. Understanding of the aerosol indirect forcing will require a combination of global observations, field measurements, and a range of modeling and analysis studies.

Global observations to determine the aerosol direct and indirect climate forcings will need to include simultaneous measurements of reflected solar and emitted thermal radiation fields as described above. The instruments measuring these two radiation fields must look at the same area at essentially the same time. Such a mission concept has been defined (Hansen et al., 1992) and recent reassessments indicate that it could be achieved at a cost of about $100M if carried out by the private sector without a requirement for undue government review panels (B. Cairns, private communication, 2011).

**Acknowledgments.** We thank Gokhan Danabasoglu, Tom Delworth, and Jonathan Gregory for providing data and papers defining responses of the NCAR, GFDL and Hadley climate models, Greg Johnson, John Lyman and Sarah Purkey for providing ocean heat data and preprints of their papers, Sarah Purkey for computing the sea level rise due to Southern Ocean warming excluding overlap with Argo data, Catia Domingues for illuminating information, Norm Loeb for references and discussion on satellite radiation budget measurements, Eric Rignot for providing data on Greenland and Antarctic ice sheet mass balance, and Robert Gibson, Wayne Hamilton, John Marshall, Don McKensie, Gregory Monahan, Tim Palmer, Bill Rossow, Ken Schatten, Gavin Schmidt, Steve Schwartz and Michael Wright for comments on a draft of this paper.